\documentclass[aps,prd,twocolumn,10pt,superscriptaddress,nofootinbib,nobibnotes,nolongbibliography]{revtex4-2}

\usepackage{amssymb}
\usepackage{graphicx}
\usepackage{amsmath}
\usepackage{hyperref}
\usepackage{multirow}
\usepackage{booktabs}
\usepackage{setspace}
\usepackage{verbatim}
\usepackage{float}
\usepackage{color}
\usepackage{ulem}
\usepackage{arydshln}
\usepackage[utf8]{inputenc}
\usepackage[table,xcdraw]{xcolor}
\usepackage{subcaption}
\usepackage{lipsum}
\usepackage{physics}

\graphicspath{{./figures/}}

\newcommand{\md}{\mathrm{d}}
\newcommand{\transp}{\mathsf{T}}

\begin{document}

\title{The MeerKAT Thousand-Pulsar Polarization Array I: \\ Properties of the Polarization and Rotation Measure Time Series Data}

\author{Michael Sarkis}
\affiliation{Department of Physics, Stellenbosch University, Matieland 7602, South Africa}
\author{Zi-Yan Yuwen}
\affiliation{Department of Physics, Stellenbosch University, Matieland 7602, South Africa}
\affiliation{Asia Pacific Center for Theoretical Physics (APCTP), Pohang 37673, Korea}
\affiliation{Institute of Theoretical Physics, Chinese Academy of Sciences (CAS), Beijing 100190, China}
\author{Yin-Zhe Ma}
\email{Corresponding author: Y.-Z. Ma, mayinzhe@sun.ac.za}
\affiliation{Department of Physics, Stellenbosch University, Matieland 7602, South Africa}
\author{Tao Liu}
\affiliation{Department of Physics and Jockey Club Institute for Advanced Study, The Hong Kong University of Science and Technology, Hong Kong S.A.R., China}
\author{Jing Ren}
\affiliation{Institute of High Energy Physics, Chinese Academy of Sciences, Beijing 100049, China}
\affiliation{Center for High Energy Physics, Peking University, Beijing 100871, China}
\author{Patrick Weltevrede}
\affiliation{Jodrell Bank Centre for Astrophysics, Department of Physics and Astronomy, University of Manchester, Manchester M13 9PL, UK}
\author{Xiao Xue}
\affiliation{Institut de Física d’Altes Energies (IFAE), The Barcelona Institute of Science and Technology, Campus UAB, 08193 Bellaterra (Barcelona), Spain}

\date{\today}

\begin{abstract}
    The polarimetry of recent pulsar observations has provided a wealth of observational data with which to test physical theories of emission mechanisms, radiative transfer and even theories that extend beyond the Standard Model. In this work, we have outlined the data analysis of the polarisation time series data of a population of 513 pulsars from the Thousand Pulsar Array observing programme, laying the foundation
for building the
MeerKAT Thousand-Pulsar Polarization Array as a probe for ultralight Axion-Like Dark Matter (ALDM). From this large dataset we have focused on the temporal trends in the observed polarisation angle (PA) through a measure we call the PA offset, and characterised the trends due to the effects of Faraday Rotation within the interstellar medium and the Earth's ionosphere, 
    as well as generic white and red noise models that are estimated within a Bayesian MCMC analysis. Then, motivated by potential extra contributions to the rotation of the PA that may not be Faraday-like, arising from the proposed ALDM field, we have investigated a derived time dependence for the rotation measure (RM) required to explain the observed PA offset.     Comparison of these estimates to RM values that are measured in typical pulsar studies, through a technique known as RM Synthesis, provides a probe of any wavelength-independent contribution to the rotation of the PA. Although we find no evidence for oscillatory behaviour within our dataset's observation timespan, 
    we do find cases of deviation from the usual RM values in several `pulsars of interest', as well as long-term linear trends in the time evolution of Faraday rotation that have been presented in the literature before.       
\end{abstract}

\maketitle

\section{Introduction}

Pulsars have proved to be valuable sources of astrophysical data over the previous several decades, with currently more than 4000 individual pulsars listed on the Australia Telescope National Facility (ATNF) Pulsar Catalogue~\cite{manchesterATNFPulsarCatalogue2005} used for a huge range of scientific goals. An important feature of pulsar emissions that is often observed is a high degree of linear polarisation, and sometimes also circular polarisation. These polarimetry observations allow for further study into the emission mechanisms of the neutron stars, especially regarding details about the pulsar's orientation, magnetosphere and emission height, with several theoretical interpretations of the data guiding the data analysis. Some historically notable works in this regard include the proposal that the Polarisation Angle (PA) is oriented by the field lines in the polar regions of the magnetosphere~\cite{radhakrishnanMagneticPolesPolarization1969,komesaroffPossibleMechanismPulsar1970} (with its relativistic counterpart~\cite{blaskiewiczRelativisticModelPulsar1991,lyutikovRelativisticRotatingVector2016}), the modelling of a `gap' between the pulsar surface and point of emission~\cite{rudermanTheoryPulsarsPolar1975} and the existence of orthogonal polarisation modes (OPMs)~\cite{backerOrthogonalModeEmission1976}.  

According to current understanding, the PA of the emission that leaves the pulsar is set by the magnetosphere. However, this initial value will change during propagation through the ionised intersteller medium (IISM) that lies between the pulsar and the Earth, mainly due to the well-known effect of Faraday Rotation (FR), with the associated characteristic Rotation Measure (RM) defining the overall magnitude of this rotation. This effect has been used to study the structure of the Galactic magnetic field (see~\cite{hanPulsarRotationMeasures2006,noutsosNewPulsarRotation2008,noutsosPulsarPolarisation2002015,hanPulsarRotationMeasures2018,setaMagneticFieldsMilky2021,dhakalProbingMagnetoionicMedium2025,oswaldThousandPulsarArrayProgrammeMeerKAT2025}), and various studies have been undertaken to characterise the temporal variation in both RM and the related Dispersion Measure quantities in pulsar populations~\cite{phillipsTimeVariabilityPulsar1991,yanRotationMeasureVariations2011,petroffDispersionMeasureVariations2013,Keith:2024kpo,kumarMultifrequencyCensus1002025}.

Recently, pulsar polarisation has also been suggested as a tool for probing the existence of ultralight models of Axion-Like Dark Matter (ALDM)~\cite{Liu:2019brz}, with several studies following this~\cite{Caputo:2019tms,Liu:2021zlt,castilloSearchingDarkmatterWaves2022,PPTA:2024mgh,Porayko2024}. Among them, the development of the concept of Pulsar Polarization Arrays (PPA)~\cite{Liu:2021zlt} is key which proposed to search for the wave signals of the ALDM in PA, characterized as sinusoidal variations over spacetime, by cross-correlating pulsar polarization data (for the generalization of this strategy to the ALDM detection through pulsar timing residuals~\cite{khmelnitskyPulsarTimingSignal2014}, see~\cite{Luu:2023rgg}). Here the signal timescale is tied to the ALDM mass by $T = 2\pi/m_{a}$, while the spatial scale is dictated by the ALDM de Broglie wavelength, $L = 2\pi/m_{a} v$. For masses characteristic of `fuzzy' dark matter ($m_a \sim 10^{-22}\;\mathrm{eV}$), these oscillations would occur on timescales of approximately 1 year and spatial scales of several hundred parsecs. The first real-data analysis, utilizing polarization data from the Parkes Pulsar-Timing-Array (PTA) program, has already provided the leading constraints on the Chern-Simons coupling of `fuzzy' ALDM~\cite{PPTA:2024mgh}. Another crucial aspect to this signal, which differentiates it further from other astrophysical sources of PA rotation like FR, is that it is completely wavelength-independent. This fact is part of the motivation for the presented analysis in this work, since in order to correctly distinguish any contributions to the polarisation of pulsar emissions from potential ALDM models and maximize the chance for a discovery on the PPA, the PA and RM time-series of the studied pulsars need to be well-characterised and understood. The study of utilizing the MeerKAT TPPA to detect ultralight ALDM and constrain its parameter space, along with its key results, will be presented in a companion paper~\cite{Yuwen:2025}.

In this work, we attempt to further characterise the properties of the PA and RM time-series data for a large sample of pulsars observed as part of the Thousand Pulsar Array (TPA) programme~\cite{Johnston:2020qxo}, laying the foundation for building the first regular-pulsar (non-MSP) PPA - the MeerKAT Thousand-PPA (TPPA). We do this primarily by defining a quantity that we call the PA offset, which is an estimator of the temporal variability in the PA from epoch-to-epoch, and studying its behaviour across the population. This definition not only provides us with a technique to search for signatures of exotic physics like ALDM, but also to reveal any potential non-FR like effects in the time-series data. 

The structure of this paper is as follows: in Section~\ref{sec:data} we describe the observational dataset used in this work, and how the PA offset values are found. In Section~\ref{sec:characterisation}, we outline the possible astrophysical processes that may have an impact on the polarisation, and characterise these according to their deterministic or stochastic properties, and in Section~\ref{sec:RM} we introduce a method for analysing the $\mathrm{RM}$ to determine the presence of any wavelength-independent PA rotation. Finally, in Section~\ref{sec:results} we present the results of the analysis and discuss possible astrophysical interpretations. 

\section{Data}\label{sec:data}

Our dataset is comprised of 1237 unique pulsars that have been observed by the TPA programme, which started off as part of the umbrella MeerKAT Key Science Project MeerTime~\cite{bailesMeerTimeMeerKATKey2018}, but is now a stand-alone extra-large programme. The observations have taken place since February 2019, and one of the programmes initial objectives was to monitor a large number of pulsars with enough signal-to-noise ratio to be able to determine accurate polarisation properties, such as RM and DM. The``fold-mode" data has been initially processed through the MeerTime pipeline~\cite{bailesMeerTimeMeerKATKey2018}, followed by the TPA-specific pipeline as described in \cite{bwk+24} which is designed to monitor pulse shape variability.

To find PA offset values, we start with standard Faraday derotation of the observed Stokes parameters, using the software package {\texttt PSRCHIVE} \cite{hvm04}, using a fixed RM (denoted $\mathrm{RM_{fixed}}$) as defined for each pulsar by the TPA programme.
This derotation is defined by the expected PA offset caused by Faraday rotation and is given by $\Delta \mathrm{PA} = \mathrm{RM_{fixed}} \lambda^2$. This offset is then used to derotate the $Q$ and $U$ parameters, using the equation $\mathrm{PA} = (1/2)\arctan(U/Q)$. For each observation epoch and each rotational phase bin the $U$ and $Q$ parameters are averaged over all frequency channels. After Faraday rotation as quantified by $\mathrm{RM_{fixed}}$ has been removed, and frequency averaging has taken place, a pulse profile (for all Stokes parameters) is obtained for each epoch. It is from this profile any possible epoch-to-epoch variability of the PA, associated with ALDM, the ISM, or intrinsic to the pulsar, is to be extracted.

The PA variability was extracted using the {\texttt{PSRSALSA}\footnote{https://github.com/weltevrede/psrsalsa} software package \cite{wel16}. To extract this PA variability, it is desirable to extract a single PA value for each epoch. However, this is complicated by the fact that the PA for different rotational phase bins are different. To account for this pulsar-specific characteristic, the median polarised profile shape is determined (see \cite{bwk+24} for details) from the individual epochs. From this median profile the characteristic PA swing of the pulsar is determined, which is then used as a template to derotate $Q$ and $U$ for each epoch, resulting in profiles which have an approximately constant $U/Q$ as function of rotational phase bin. 

Therefore, to arrive at a single data point for each epoch, the $Q$ and $U$ values are averaged over the pulse longitude bins that contain the signal, i.e. in the `on' bins where the pulse is located (see \cite{bwk+24} for details of how the on bins were determined). This single set of frequency- and longitude-averaged Stokes parameters are then used to determine the PA, which is used in the analysis as a time-series of PA offset values. 

Each datapoint for PA offset has a corresponding error estimate based on the observed RMS of the $Q$ and $U$ parameters in off-pulse regions, which are propagated through the above equations, i.e. 
\begin{align}\label{eqn:PA_error}
    \sigma_\mathrm{PA} = \frac{1}{2}\frac{\sqrt{U^2\sigma_Q^2+Q^2\sigma_U^2}}{Q^2+U^2} \,.
\end{align}

\subsection{Data cleaning and preparation}

Although the full TPA observational catalogue consists of 1237 unique pulsars, the observing cadence of each pulsar is not equal. In particular, the overall TPA sample was observed at least once with a high sensitivity observation utilising the full MeerKAT array, with a smaller subsample that is being monitored regularly using two sub-arrays \cite{swk+21}. 
Because of changes in the allocated telescope time, and revisions of the monitoring scheduling, there are three broadly defined observation `groups', with pulsars of less than 20 observing epochs, those of between 40 and 60 epochs, and those above 60 epochs (these groups are shown in Fig.~\ref{fig:Nepochs}). To ensure the statistical power of the time-series analysis we perform is valid, we currently only consider pulsars that have more than 20 valid data points. Before this criteria is considered, we also perform a simple outlier removal for the $p$-th pulsar and $n$-th individual observing epoch using the following two conditions:
\begin{align}\label{eqn:datacuts}
    |\Delta\mathrm{PA}_{p,n} - \mathrm{med}(\Delta\mathrm{PA}_{p})| &> 4~\mathrm{std}(\Delta\mathrm{PA}_{p})\\[0.5em]
    |\sigma_{p,n} - \mathrm{med}(\sigma_{p})| &> 4~\mathrm{std}(\sigma_{p})~, 
    \label{eqn:datacuts2}
\end{align}
where $\mathrm{med}$ and $\mathrm{std}$ denote the median value and standard deviation of the given pulsar's full dataset, and $\sigma_{p,n}$ is the uncertainty associated with each PA offset. Finally, if we find that our computed estimates of the ionospheric FR and/or RMS are missing (i.e. there is an unknown value during that observing session) because of the unavailability of TEC map information (see Section~\ref{sec:ionosphere}), we also cut that data point from the rest of the analysis. The above data cleaning procedure leaves the majority of pulsars' data unaffected, with only 451 of the $\sim 32$k total number of data points (across all pulsars) cut. An instructive visualisation of this datacut procedure for the pulsar J2039$-$3816 is provided in Fig.~\ref{fig:example_datacut}. Both rejected points correspond to observations badly affected by radio frequency interference.

With the above conditions, the number of pulsars considered for further analysis drops from the initial 1237 to 513 (i.e. 724 pulsars are dropped), which we refer to later as our full population. In the resulting dataset, the mean number of observing epochs is $\approx 60$, with a mean average cadence of $\sim 35$ days, and the median and standard deviation of the PA offset uncertainties are $\mathrm{med}(\sigma_n) = 1.41~\mathrm{deg}$ and $\mathrm{std}(\sigma_n) = 0.52~\mathrm{deg}$, respectively. 

\begin{figure}[htb]
    \centering
    \includegraphics[width=0.95\columnwidth]{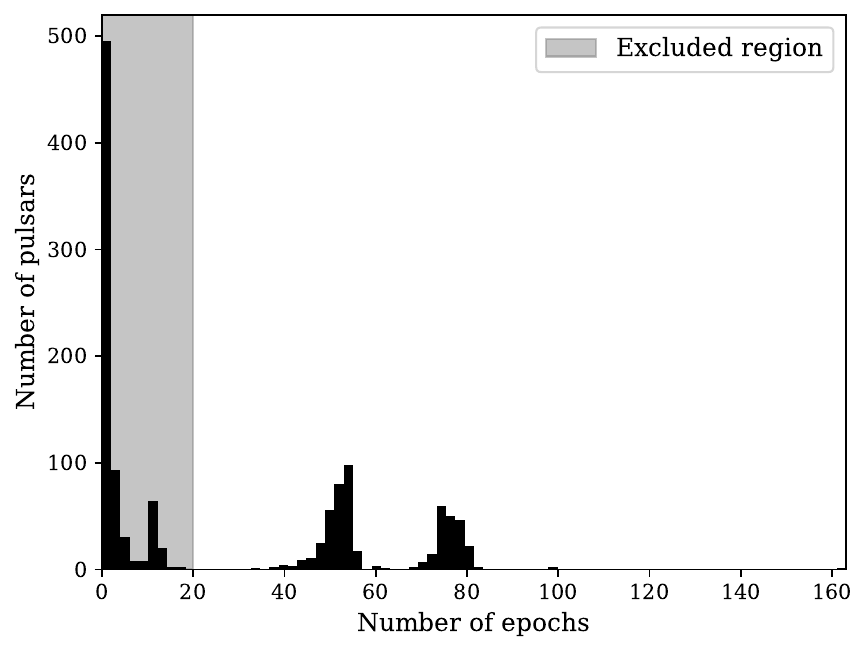}
    \caption{Number of observation epochs for each of the 1237 pulsars in the population, after outlier removal and data cleaning. Pulsars with less than 20 unique epochs are excluded from our analysis.}
    \label{fig:Nepochs}
\end{figure}

\begin{figure}[htb]
    \centering
    \includegraphics[width=0.95\columnwidth]{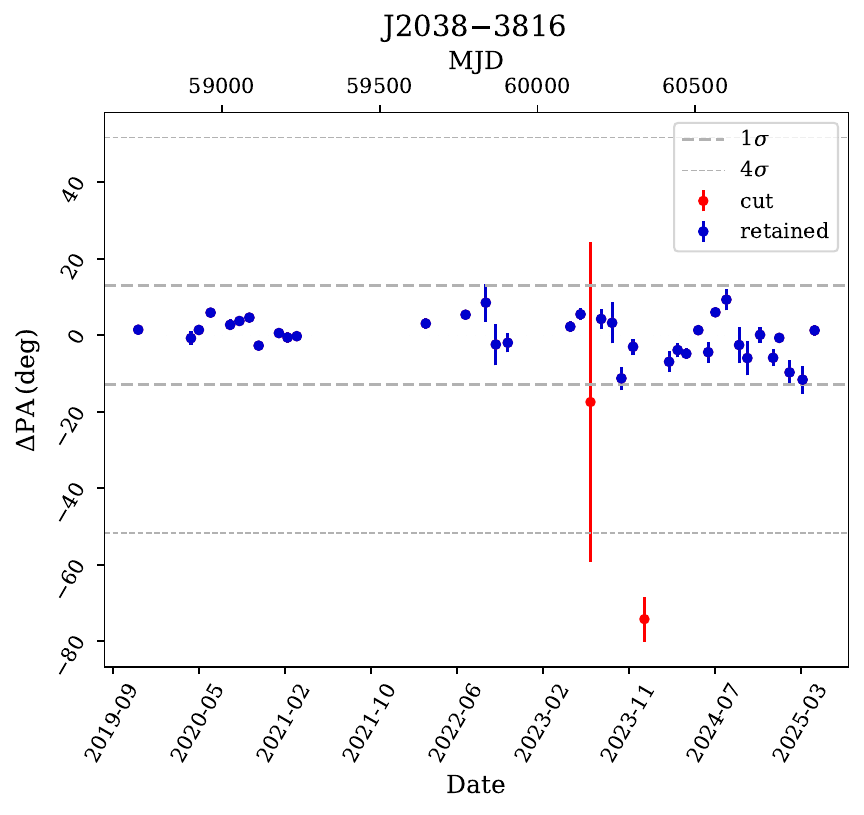}
    \caption{Example of derived PA offset data from pulsar J2038$-$3816, and a visualisation of the cutting of outlier data points. In this example, we can see examples of data points that satisfy both of our datacut criteria, having both an outlying PA offset value as well as a value with extremely large uncertainty.}
    \label{fig:example_datacut}
\end{figure}

\section{Characterisation of polarisation time-series data}\label{sec:characterisation}


When analysing the polarisation time-series data for both the PA residual and RM values, we consider both a set of effects from expected astrophysical sources and generic noise terms. When modelling these effects, we further split the astrophysical sources of RM that impact each pulsar's emission into two regions, the IISM and the ionosphere of the Earth. In these regions, we expect rotation of the PA to be due FR, which is defined as 
\begin{equation}\label{eqn:FR}
\mathrm{FR} = \mathrm{RM} \lambda^2 \,
\end{equation} 
where $\mathrm{RM}$ is defined as
\begin{align}\label{eqn:RM}
    \mathrm{RM} &= \frac{e^3}{2\pi m_{\rm e}^2c^4} \int_0^d \,n_{\rm e}(s) B_{\parallel}(s)\, \dd{s} \,\\
    \left( \frac{\mathrm{RM}}{\mathrm{rad\,m^{-2}}}\right)&\approx 0.81\int_0^d \left(\frac{n_{\rm e}(s)}{\mathrm{cm^{-3}}}\right)\left(\frac{B_{\parallel}}{\mathrm{\mu G}}\right) \dd{s} 
\end{align}
where $n_{\rm e}(s)$ and $B_{\parallel}(s)$ are the electron density and magnetic field strength along the line of sight $s$ to the pulsar located at distance $d$, respectively, and $e, m_{\rm e}$ are the electron charge and mass.

\subsection{Ionospheric effects}\label{sec:ionosphere}

The ionosphere of the Earth is a layer of free electrons and ionised molecules tens to hundreds of kilometers high in the Earth's atmosphere. Since the ionisation of this atmospheric layer is caused by solar radiation, changes in the solar irradiance within day-night, seasonal and solar activity cycles introduce a significant variability to the ion density of this layer, and hence the amount of FR experienced by EM radiation propagating through it. Recent investigations into the variability of the ionosphere and its effect on pulsar measurements have been made~\cite{sotomayor-beltran:2013vma,poraykoTestingAccuracyIonospheric2019,poraykoValidationGlobalIonospheric2023} which suggest that the ionosphere is likely to be the dominant component of temporal variability of RM. The RM due to the ionosphere is usually modelled and accounted for using one of a variety of tools that combine geomagnetic models and global TEC (Total Electron Content) maps of the ion density obtained from Global Navigation Satellite System (GNSS) data. While there does not seem to be a current catch-all solution for precisely accounting for these effects in pulsar data, due primarily to the complex nature of fluctuations in the ionosphere density, one generally sees a significant reduction in seasonal and diurnal effects that can be attributed to the ionosphere by using these models. 

In this work we employ two public codes, \texttt{spinifex}~\cite{mevius_2025_15000430} and \texttt{ionFR}~\cite{sotomayor-beltran:2013vma}, to evaluate the ionospheric component, and apply the corrections provided by these codes directly to our observed PA residuals. In the case of \texttt{spinifex}, we use the Single Layer Model (SLM) with the IONEX-based models provided by Technical University of Catalonia's (UPC's) Chapman server~\footnote{\url{https://chapman.upc.es/tomion/rapid/}}, and more specifically we use the provided UQRG maps~\cite{orusImprovementGlobalIonospheric2005}. In the case of \texttt{ionFR}, we follow~\cite{Keith:2024kpo} and use the TEC maps obtained from NASA's CDDIS~\footnote{\url{https://cddis.nasa.gov/archive/gps/products/ionex/}}. We note that for some observation epochs, TEC maps from the above CDDIS source we unavailable, which led to a reduction in the total number of viable data points to be used in the full analysis. When using the~\texttt{ionFR} code, we found a total of 415 total pulsars that survived the data selection criteria, compared to the full number of 513 that was considered when using~\texttt{spinifex}. Although a detailed study of the effectiveness of various GNSS data and TEC maps on our pulsar population is outside the scope of this work, we consider the above two codes and sets of TEC maps mainly for comparative purposes.
We also note that the average cadence of all 513 pulsar observations is $\sim 35$ days, which allows for a good sampling rate of the seasonal ionospheric variations but not diurnal effects.
We therefore use $\delta\mathrm{PA} \equiv \Delta {\rm PA}^{\rm obs} - \Delta \mathrm{PA^{ion}}$ to denote the {\it ionosphere-corrected observed data}, which is the actual data input we use for MCMC and parameter constraints.


\subsection{IISM effects}\label{sec:ISM}

As the pulsar emission travels through the IISM, the FR affecting the PA residuals can be characterised by the usual RM relationship for pulsar emission given by Eq.~\eqref{eqn:RM}. Thus, variations in either the free electron density or the magnetic field strength along the line of sight may alter the observed PA at different epochs. After analysing the behaviour of the RM time series of 597 pulsars in the TPA, Ref.~\cite{Keith:2024kpo} found that the `RM slope' (i.e. the measured gradient in the RM time series, in $\mathrm{rad\, m^{-2}\, yr^{-1}}$) was more than $3\sigma$ away from zero in 58 pulsars, which suggests a potential long-term variability in the RM of these pulsars. When compared to a similar analysis of the Dispersion Measure (DM) data, the authors suggest that this long-term variability may be caused by two phenomena: the turbulence in the IISM (which results in equal RM and DM effects) and the presence of discrete magnetised structures along the line of sight, such as supernova remnants (which results in more RM than DM). To model this potential long-term variability, we take a similar approach to~\cite{PPTA:2024mgh} and adopt a simple second-order polynomial in time, i.e. 

\begin{equation}\label{eqn:det_polynomial}
    \Delta \mathrm{PA^{poly}} = \psi^{(0)} + \psi^{(1)} \left(\frac{t}{T_{\mathrm{yr}}}\right) + \psi^{(2)} \left(\frac{t}{T_{\mathrm{yr}}}\right)^2~,
\end{equation}
where $T_{\mathrm{yr}}$ is a scaling parameter which represents a time scale of 1 year. We also take into account the annual variation in the line-of-sight of the pulsar through the IISM due to the motion of the Earth's solar orbit, which we model as a harmonic function with annual period, i.e.

\begin{equation}\label{eqn:det_annual}
    \Delta \mathrm{PA^{yr}} = \psi^{(s)} \sin\left[2\pi\left(\frac{t}{T_{\mathrm{yr}}}\right)\right] + \psi^{(c)} \cos\left[2\pi\left(\frac{t}{T_{\mathrm{yr}}}\right)\right] \,.
\end{equation}

The effect of Eq.~\eqref{eqn:det_annual} is similar to the one described in Ref.~\cite{Keith:2012ht}, in which variations of a pulsar's line-of-sight through regions of differing IISM plasma density, due to the Earth's yearly orbit, is attributed to the observed annual modulation in some pulsars' DM time series data. This effect is expected to be highly dependent on pulsar position and the presence of a significant plasma density gradient in the line-of-sight, with pulsars that have higher parallaxes more susceptible to the effect in general. We model the IISM effect as a data vector $\Delta {\rm PA}^{\rm IISM} = \Delta \mathrm{PA^{poly}} + \Delta \mathrm{PA^{yr}}$ for each pulsar, which can be obtained from the corresponding epochs at observation time $t_{n}$ as $\Delta \mathrm{PA}^\mathrm{IISM} = M \psi$ with
\begin{align}
    &M = \left(
    \begin{array}{ccccc}
        1 & \tilde{t}_{1} & \tilde{t}_{1}^2 & \sin\tilde{t}_{1} & \cos\tilde{t}_{1} \\[0.5em]
        1 & \tilde{t}_{2} & \tilde{t}_{2}^2 & \sin\tilde{t}_{2} & \cos\tilde{t}_{2}  \\[0.5em]
        \vdots & \vdots & \vdots & \vdots & \vdots \\[0.5em]
        1 & \tilde{t}_{N} & \tilde{t}_{N}^2 & \sin\tilde{t}_{N} & \cos\tilde{t}_{N}
    \end{array} \right)~, \\
    &\psi^\transp = \left( \psi^{(0)},\psi^{(1)},\psi^{(2)},\psi^{(s)},\psi^{(c)} \right)~,
\end{align}
where $\tilde{t} \equiv t/T_\mathrm{yr}$.

\subsection{Expected noise components}\label{sec:noise}

Outside of the ionospheric and IISM effects discussed above, we attempt to characterise the stochastic noise in our data with a white noise term, $\Delta\mathrm{PA}^\mathrm{w}$, describing temporally uncorrelated noise, and red noise, $\Delta\mathrm{PA}^\mathrm{r}$, accounting for any temporally correlated noise. This description of noise is similar to a basic noise model in gravitational wave searches performed with PTAs, and has also been employed in Ref.~\cite{PPTA:2024mgh}. The covariance matrix for the white noise is given by
\begin{align}\label{eq:fullwn}
    C^\mathrm{w}_{n,m} = \left( \left(\mathrm{EFAC}~\sigma_{n}^\mathrm{c}\right)^2 + \mathrm{EQUAD}^2 \right)\delta_{m,n} \, ,
\end{align}
where $\sigma_{n}^\mathrm{c}$ is the ionosphere corrected error of the PA offset for the $n$-th observation epoch, defined as 
\begin{align}
    \sigma_n^\mathrm{c} = \sqrt{\sigma_n^2 + \sigma_\mathrm{RM}^2 \lambda^4}\, ,
\end{align}
where $\sigma_\mathrm{RM}$ is the RM error obtained by the given ionosphere model, either \texttt{spinifex} or \texttt{ionFR}. The parameter pair ($\mathrm{EFAC}$, $\mathrm{EQUAD}$) stands for ``Error FACtor'' and ``Error added in QUADrature'' that are often used in pulsar timing studies~\cite{Lentati:2013rla}. EFAC scales the observational errors $\sigma_{n}$ due to any potential mis-calibration, and should be of order unity if the observational errors are estimated accurately. EQUAD describes other potential white noise components not accounted for in the observation error but intrinsic to the pulsar~\cite{Nice:1995,Shannon:2014kha}, such as pulse ``jitter"~\cite{Liu:2011te}. Note that in the definition above the EQUAD parameter is not rescaled by EFAC, which is slightly different from the convention in Ref.~\cite{NANOGrav:2023ctt}. Given that our polarisation residual time-series data is found for a single frequency band, we do not include the off-diagonal white noise term, which is usually modelled with block diagonal matrices and the ECORR parameter. 

The red noise covariance matrix is given as follows,
\begin{equation} \label{eq: Cred}
    C^\mathrm{r}_{n,m} = (S^{\rm r})^2\int_{f^\mathrm{L}}^{f^\mathrm{H}} \frac{\md f}{f_\mathrm{yr}} \left(\frac{f}{f_\mathrm{yr}}\right)^{\Gamma} \cos\left[2\pi f(t_{n} - t_{m})\right] \,,
\end{equation}
where $S^{\rm r}$ is the characteristic strength of red noise and $\Gamma_p$ is the power law index, $f^\mathrm{L}$ and $f^\mathrm{H}$ are the low- and high-frequency cut-offs respectively, and $f_\mathrm{yr} = 1/T_\mathrm{yr}$. Thus, the red noise add off-diagonal terms to account for correlations among different observation epochs for each pulsar. We set $f^\mathrm{L} = \Delta f= 1/T_{\mathrm{obs}}$, where $T_{\mathrm{obs}}$ is the total observation time span. The high-frequency cut-off is set to be $k_{\mathrm{max}}\Delta f$ with an integer $k_{\mathrm{max}}$. Then Eq.~\eqref{eq: Cred} can be discretized as
\begin{eqnarray} 
\label{eq: Cred summation}
C^\mathrm{r}_{n,m} = (S^{\rm r})^2 \frac{\Delta f}{f_\mathrm{yr}} \sum_{k=1}^{k_{\mathrm{max}}}  \left(\frac{k\Delta f}{f_\mathrm{yr}}\right)^{\Gamma}\cos\left[2\pi k\Delta f(t_{n} - t_{m})\right]. \nonumber \\
\end{eqnarray}

\subsection{Bayesian likelihood and parameter inference}\label{sec:likelihoods}

The estimation of individual noise parameters is carried out with a straightforward Bayesian likelihood analysis with the above generic white and red noise definitions, and by using the PTMCMC sampler code\footnote{https://doi.org/10.5281/zenodo.1037579}. For the four noise parameters $\{\mathrm{EFAC}, \mathrm{EQUAD}, S^{\rm r}, \Gamma\}$ of each pulsar, we set the prior probability distributions as follows:

\begin{align}
    {\rm Pr}(\log_{10}(\mathrm{EFAC}))                 &= U[-2,2] \nonumber \\
    {\rm Pr}(\log_{10}(\mathrm{EQUAD}/\mathrm{rad}))   &= U[-8,2] \nonumber \\
    {\rm Pr}(\log_{10}(S^{\rm r}/\mathrm{rad}))        &= U[-8,2] \nonumber \\
    {\rm Pr}(\Gamma) &= U[-8,0] 
\end{align}
where $U$ is a uniform distribution. The upper bound for $\Gamma$ is set to be $0$ to allow for mathematical convergence of the integral in the covariance matrix of red noise Eq.~\eqref{eq: Cred summation}, and a negative $\Gamma$ ensures no dependence on the high-frequency cut-off $f^{\mathrm{H}}$ for large enough values of $f^{\mathrm{H}}$. For the parameters in Eqs.~\eqref{eqn:det_polynomial} and~\eqref{eqn:det_annual}, we set their priors with the uniform distribution, i.e. $\mathrm{Pr}(\psi^{(i)}/\mathrm{rad}) = \mathrm{U}[-2,2]$ for $\psi^{(i)} \in \{\psi^{(0)},\psi^{(1)},\psi^{(2)},\psi^{(c)},\psi^{(s)}\}$. In most results presented here the coefficients $\psi^{(i)}$ are treated as nuisance parameters and marginalised, however in Appendix~\ref{sec:marginalisation} we also show their posteriors from MCMC sampling as free parameters. 


We now combine the data vector $(\delta \rm PA-\Delta {\rm PA}^{\rm IISM})\equiv (\delta {\rm PA}-M\psi)$ and its covariance matrix $C$ to formulate the log-likelihood
\begin{align}
\label{eq: lnL}
    \ln \mathcal{L} = -\frac{1}{2}\left(\delta \mathrm{PA}- M\psi\right)^{\transp} C^{-1} \left(\delta \mathrm{PA}- M\psi\right)  - \frac{1}{2}|2\pi C| ,
\end{align}
where the noise covariance matrix is a combination of white and red noise terms, i.e. $C = C^{\mathrm{w}} + C^{\mathrm{r}}$.

By considering the parameters in Eqs.~\eqref{eqn:det_polynomial} and~\eqref{eqn:det_annual} as nuisance parameters, the $\psi^{(i)}$ can be analytically integrated to obtain a marginalised likelihood function in a similar way to~\cite{vanhaasterenMeasuringGravitationalwaveBackground2009,PPTA:2024mgh}, whose details are given in Appendix.~\ref{sec:marginalisation}, resulting in a marginalised likelihood as follows,
\begin{align}\label{eq: lnLm}
    \ln \mathcal{L}_\mathrm{m} &=  \frac{1}{2}(\delta\mathrm{PA}^c)^\transp C_M^{-1} \delta\mathrm{PA}^c  - \frac{1}{2}\ln |2\pi C_M| \\
    &\quad~ - \frac{1}{2}\left(\delta \mathrm{PA}\right)^\transp C^{-1} \delta\mathrm{PA} - \frac{1}{2}\ln |2\pi C| + \mathrm{const.} \nonumber
\end{align}
where $C_M=M^\transp C^{-1} M$ and $\delta\mathrm{PA}^c=M^\transp C^{-1}\delta\mathrm{PA}$. A constant term in $\ln \mathcal{L}_\mathrm{m}$ does not affect the result and can be removed by performing a normalization, therefore the last constant term in Eq.~\eqref{eq: lnLm} can be neglected.

When evaluating the performance of each model comparatively, computing the Bayes Factor by directly integrating the marginal likelihoods of each model is computationally expensive because of large amount of free parameters in the model. Instead, we make use of the Product Space method~\cite{lodewyckxTutorialBayesFactor2011}, to determine the significance of red noise in the analysis, in a similar manner to Ref.~\cite{PPTA:2024mgh}. The product-space method is a transdimensional scheme that uses a Cartesian product of the parameter spaces from the models under comparison to create one single mixture model. Within this mixture model, a hyper-parameter $n$ is introduced that can be used in the following new conditional likelihood:
\begin{align}
    \mathcal{L}(n,\vartheta|d) = \left\{\begin{array}{lc}
        \mathcal{L}_{\mathrm{model1}}(\vartheta|d) ~,&-1\leq n <0~, \\
        \mathcal{L}_{\mathrm{model2}}(\vartheta|d) ~,&0\leq n < 1 ~. 
    \end{array}
    \right.
    \label{eq: likelihood hyper-parameter}
\end{align}
where $\vartheta$ is the Cartesian product of the parameter sets for each model, i.e. $\vartheta = \vartheta_1 \otimes \vartheta_2$, and $d$ denotes the pulsar data. Using this hybrid likelihood describing a mixture of the two models, the Bayes Factor ratio can be estimated from the ratio of volume fraction occupied by each model in the posterior. From a computational perspective, the Bayes Factor ratio is then estimated by sampling between each likelihood according to the scheme above, and taking the ratio of number of draws from each model to the total number of draws, i.e.
\begin{align}
    \mathrm{BF}_{\mathrm{model1}}^{\mathrm{model2}} = \frac{Z_\mathrm{model2}}{Z_\mathrm{model1}} \approx \frac{N_{\mathrm{model2}}}{N_{\mathrm{model1}}}~\,,
\end{align}
where $Z$ denotes the marginal likelihood or Bayesian evidence for each model under consideration, and $N$ is the number of sample draws from the posterior distribution of each model.~\footnote{Note that the Bayes Factor estimated from this method is constrained by the finite number of samplings in the MCMC chain. In our settings, $|\ln BF| \gtrsim 10$ suggests only a few visits to the disfavoured model, thereby the numerical magnitude of $\ln BF$ might be inaccurate. This corresponds to the case that almost all posterior samples are assigned to the preferred model, and therefore the model preference itself still remains clear.}

\subsection{RM Variability}\label{sec:RM}

The expected astrophysical contributions to observed PA values of pulsar emissions are often well-modelled by FR due to the effects described above, and are well-studied in the literature. However, there has also been some recent interest in using pulsar polarisation data for tests of other physical theories, which includes BSM theories like ALDM~\cite{Liu:2019brz,castilloSearchingDarkmatterWaves2022,Liu:2021zlt,PPTA:2024mgh,Porayko2024}. In these theories, there are potential contributions to the observed PA that are not due to the usual effects of a magnetised plasma, and instead due to the proposed birefringent dispersion relation of light traversing a ALDM field. In this scenario the PA would be modified in a way that does not correspond to the $\lambda^2$ relationship of Eq.~\eqref{eqn:FR}, and is expected to be wavelength-independent~\cite{Liu:2019brz}. For this reason, we describe an initial investigation into the behaviour of the observed RM time-series from the TPA dataset, in a similar vein to Ref.~\cite{Keith:2024kpo}, though with specific interest into any potentially non-wavelength-dependant behaviour of the RM. We note that a Generalised Rotation Measure (GRM), which allows a free powerlaw index in wavelength and has been studied for example in the case of a magnetar~\cite{lowerLinearCircularConversion2024}, as well as the more detailed `Faraday Mixing Scenario'~\cite{wangPropagationinducedFrequencydependentPolarization2025} which considers the additional effects of Faraday conversion and absorption, in the case of FRBs.  


To probe this effect in our data, we describe below two different values of RM that are obtained in separate and independent manners. 
\begin{itemize}
    \item Firstly, we consider a method that uses an algorithm known as RM Synthesis, first described in~\cite{brentjensFaradayRotationMeasure2005}, which computes the contributions of FR from different regions within a generalised spatially-extended source. In the case of pulsars, which can be considered point-like, the evaluation of the RM typically then involves optimising the linear polarization intensity of the pulsar emission for various RM values. Specifically, we use the same method as in~\cite{Keith:2024kpo} and other TPA-related works, which relies on the \texttt{PSRSALSA} software tool described in \cite{ijw+19}.
    The key factor in the determination of these RM values is that they are computed assuming a FR-like relationship with the wavelength of the EM wave. In this work, these values are labelled as ``RM synth''. 
    \item Secondly, we consider RM values that are derived from our PA offset values, which were described in Section~\ref{sec:data}. We infer the corresponding RM offsets by de-rotating the PA offsets with the geometric mean of the observing bandwidth ($\sim 1280$ MHz). These RM values would be required to explain the time variability of the PA offsets and are labelled as ``RM offset''.
\end{itemize}
To allow direct comparison with the independently derived RM synth values, the ``RM offset'' values are added to the fixed (time-independent) RM used to initially derotate the data and derive the PA offsets.
Because the PA offsets are initially obtained by a method that is completely independent of any assumed dependence on wavelength, i.e. by finding a template PA swing and derotating the Stokes parameters using this template, the RM from this technique should give us 
complementary information and comparison with the RM synth values allow us to identify any potential non-FR-like contributions to PA rotation.

Further, we investigate an aspect of the above comparison that may be of interest, which is of any temporal variation in the various RM measures. There have been several studies into the variations of RM, and the similar effect of variations in DM, of pulsar emissions~\cite{phillipsTimeVariabilityPulsar1991,petroffDispersionMeasureVariations2013,yanRotationMeasureVariations2011,Keith:2024kpo}. In these studies, there is evidence in the data of some pulsars for long-term variation of RM (or DM) in the form of a perceived linear trend over time, while other pulsars show evidence of very stable RM. In this work we can characterise our two independent methods for deriving the time-series RM synth and RM offset with two functions, say $f_1(\vec{D}(t))$ and $f_2(\vec{D}(t))$ respectively, where $\vec{D}$ is the raw Stokes data from the telescope, to obtain a new function, $f_{\mathrm{res}} = f_1 - f_2$. We assert that this residual function should be 0 if all sources of PA rotation contribute to the RM in a manner that obeys the standard FR relationship with wavelength, as in Eq.~\eqref{eqn:FR}. 

To test the variability of the RM measures, we thus consider both linear trend models and Lomb-Scargle Periodograms (LSP) for each pulsar in the dataset. The linear analysis, performed in a similar manner to~\cite{yanRotationMeasureVariations2011,petroffDispersionMeasureVariations2013,Keith:2024kpo}, should reveal any first-order evidence of long-term variability which is of a timescale longer the current observation timespan. For this we utilise the \texttt{lmfit}~\cite{newvilleLMFITNonLinearLeastSquares2025} software package, which has a suite of curve-fitting and data modelling functionality. Specifically, we assume that $f_{\mathrm{res}} = at + b$, i.e. a linear function, and we fit for the slope and intercept parameters $a$ and $b$ using the least-squares minimisation within \texttt{lmfit}. In all cases we use the default Levenberg-Marquardt optimisation algorithm, and analyse the estimated parameters -- particularly the slope ($a$) -- for deviations from 0. 

On the other hand, the LSP~\cite{vanderplasUnderstandingLombScarglePeriodogram2018} should provide evidence for any periodicity on shorter intervals. This is a standard tool for finding periodic signals within non-uniformly sampled data, which is applicable to all of our time-series data, and has been used before in analysing pulsar polarisation behaviour~\cite{castilloSearchingDarkmatterWaves2022, Porayko2024}. We have used the \texttt{LombScargle} class within the Astropy Python library~\cite{astropycollaborationAstropyProjectSustaining2022} to perform the calculations in this work, which is based on the framework laid out in~\cite{vanderplasUnderstandingLombScarglePeriodogram2018}. Within this framework, the LSP is used as an estimator of the Fourier power spectrum of the data, and is constructed by proposing sinusoidal models over a set of frequencies that are then minimised with respect to a standard $\chi^2$ statistic, which may also include noise in the data. The best-fit model from this minimisation at each frequency is then compared to some reference model to determine the LSP. This interpretation of the LSP is equivalent to an estimator that uses the discrete Fourier power spectrum that has been generalised to account for non-uniformly sampled data~\cite{vanderplasUnderstandingLombScarglePeriodogram2018}, and in both cases should capture any underlying frequency information in the time-series data. We have also computed False Alarm Probability (FAP) levels for each LSP, which is a probability of seeing a peak if there is no underlying signal (i.e. a spurious peak) and a common method for assessing the significance of each peak. In all cases we have computed the 10\% and 1\% probability levels using the bootstrap technique~\cite{vanderplasUnderstandingLombScarglePeriodogram2018} in the \texttt{LombScargle} class, using the default number of 1000 samples. 

\section{Results and Discussion}\label{sec:results}

\subsection{Noise model comparison}

\begin{table*}[ht]
    \begin{tabular}{c| c c | c c | c c | c c | c  c | c c} 
        \multirow{2}{*}{Pulsar name} & \multirow{2}{*}{$\ln\mathrm{BF}$} & \multirow{2}{*}{${}^{\rm spinifex}_{\rm ionFR}$}
        & \multicolumn{2}{c|}{${}^{{\rm 1st\mbox{-}order ~poly}}_{\rm 0th\mbox{-}order ~poly}$}
        & \multicolumn{2}{c|}{${}^{{\rm 2nd\mbox{-}order ~ poly}}_{\rm 1st\mbox{-}order ~poly}$}
        & \multicolumn{2}{c|}{${}^{{\rm 2nd~poly+harmonic}}_{\rm 2nd~poly}$}
        & \multicolumn{2}{c|}{${}^{{\rm 2nd~poly+white}}_{\rm 2nd~poly}$}
        & \multicolumn{2}{c}{${}^{{\rm white}+{\rm red}}_{\rm white}$} \\
        \cline{4-13}
        & & & spinifex & ionFR & spinifex & ionFR & spinifex & ionFR & spinifex & ionFR & spinifex & ionFR \\
        \hline
            J1935$+$1616 && 5.88 & -1.48 & -1.23 & -2.66 & 0.16 & -8.62 & 5.40  & 9.57 & 9.43 & -0.48 & -0.36 \\	
            J1645$-$0317 && 9.03 & -1.52 & -3.04 & -1.33 & -1.45 & -6.06 & -7.56 & 8.11 & 8.33 & -0.18 & -0.49 \\	
            J1913$-$0440 && 7.40 & -3.24 & -2.93 & -0.89 & -1.78 & -0.69 & -4.69 & 8.62 & 9.72 & -0.73 & -0.53 \\	
            J1600$-$5044 && 6.89 & -0.23 & 0.75  & -0.83 & -1.14 & -5.12 & -5.17 & 10.82 & 8.95 & -0.74 & -0.50 \\	
            J1932$+$1059 && 7.85 & -2.32 & -1.98 & -2.04 & -0.72 & -8.02 & -1.06 & -4.51 & 3.94 & -0.73 & -0.17 \\	
            J1709$-$4429 && 11.51& -3.60 & -3.01 & -2.07 & -1.25 & -7.32 & -2.34 & 9.21 & 9.57 & 0.34  & -0.57 \\	
            J1820$-$0427 && 6.31 & -2.66 & -2.53 & -0.93 & -1.06 & -7.40 & -6.14 & 9.32 & 7.90 & -0.76 & -0.65 \\	
            J1752$-$2806 && 7.21 & -2.50 & -2.33 & -0.90 & -0.75 & -6.12 & -6.36 & 7.80 & 7.60 & -0.47 & -0.60 \\	
            J1359$-$6038 && 7.87 & 5.51  & 1.79  & -1.22 & -0.79 & -3.35 & 3.60  & 7.80 & 6.78 & 7.60  & -0.52 \\	
            J0820$-$1350 && 6.38 & -3.42 & -2.17 & -1.48 & 1.15 & -6.81 & -6.25 & 9.03 & 9.90 & -0.75 & -0.63 \\	
        $\vdots$ &&&&&&&&&&  \\
        \hline
    \end{tabular}
    \caption{Summary table of Bayes Factor for comparing different noise modellings. The values in the second column are obtained by comparing the two ionosphere models with the harmonic terms and red noises turning on. The $3$rd to $8$th columns show the Bayes factors for comparing different IISM modellings, where ``poly'' stands for polynomials, with white and red noises included in both cases. The $9$th and $10$th column compares the full white‑noise model against a minimal white‑noise model with a quadratic IISM modelling. 
    The last two columns show the Bayes factors for including both white and red noise relative to white noise only, with harmonic terms included in both cases.}
    \label{tab:noise_compare}
\end{table*}


\begin{figure*}
    \centering
    \includegraphics[width=0.44\textwidth]{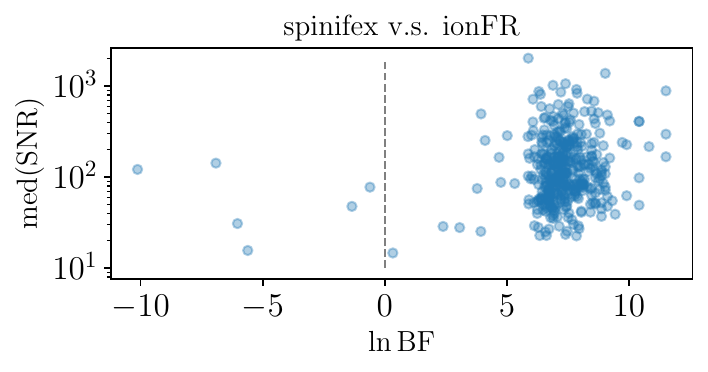} \\
    \includegraphics[width=0.4\textwidth]{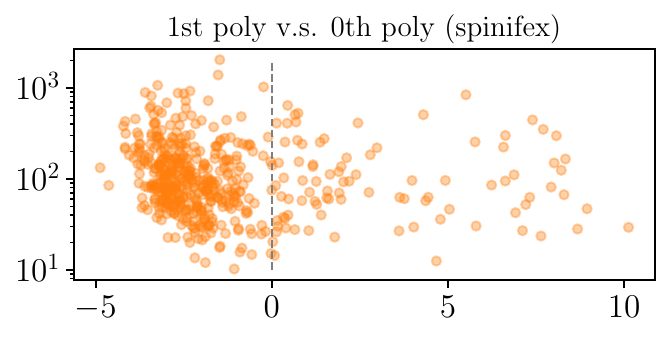}
    \includegraphics[width=0.4\textwidth]{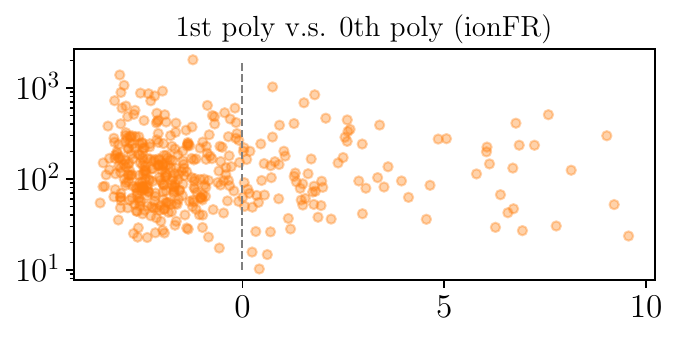} \\
    \includegraphics[width=0.4\textwidth]{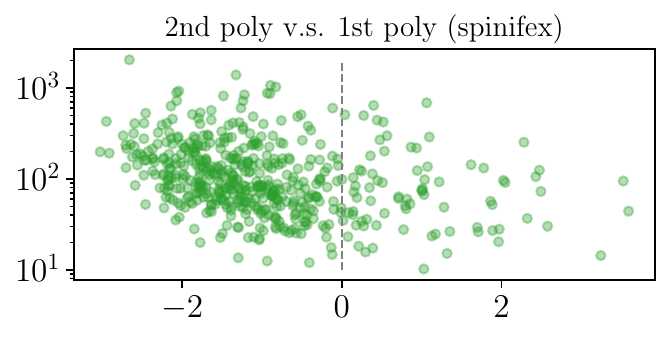}
    \includegraphics[width=0.4\textwidth]{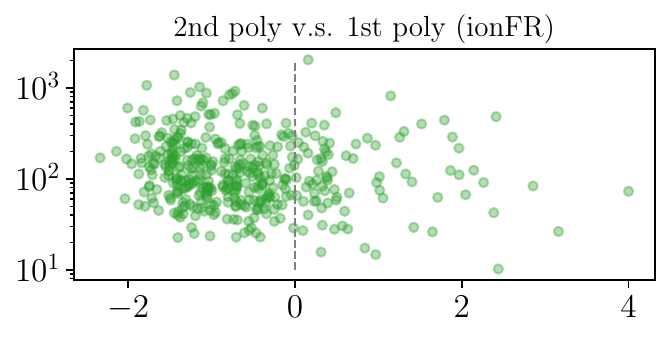} \\    \includegraphics[width=0.4\textwidth]{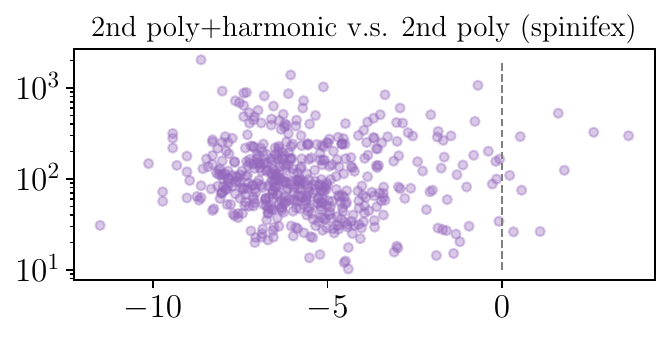}
    \includegraphics[width=0.4\textwidth]{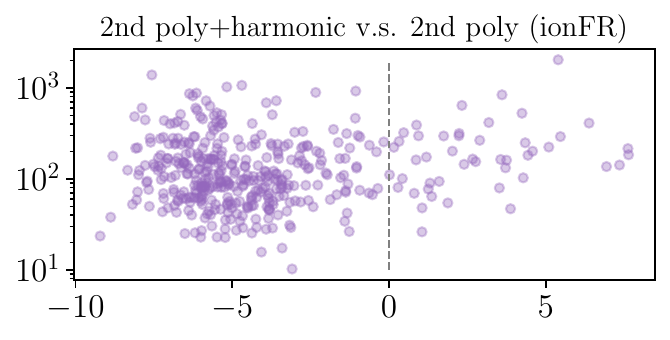} \\
    \includegraphics[width=0.4\textwidth]{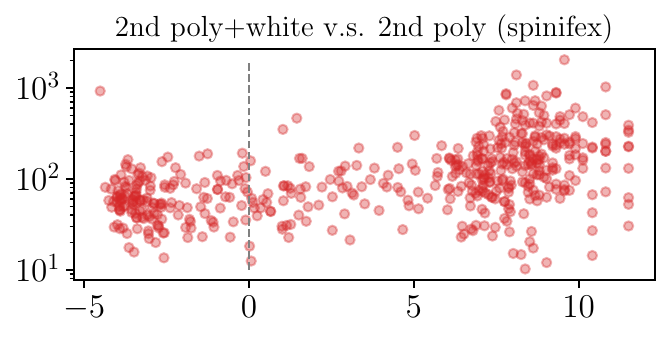}
    \includegraphics[width=0.4\textwidth]{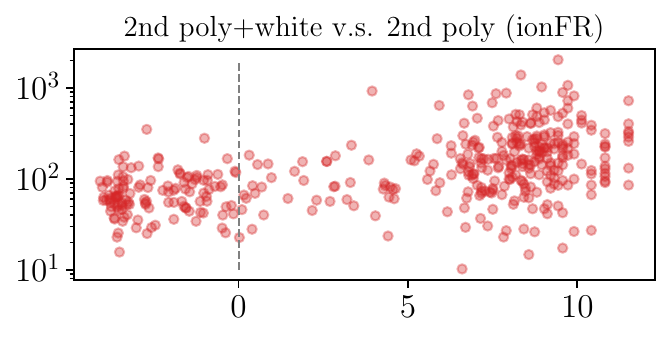} \\

    \includegraphics[width=0.4\textwidth]{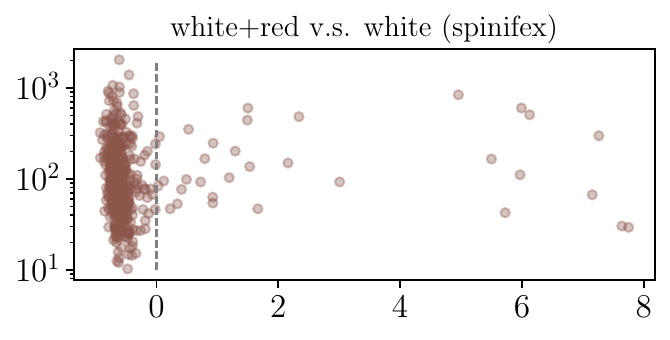}
    \includegraphics[width=0.4\textwidth]{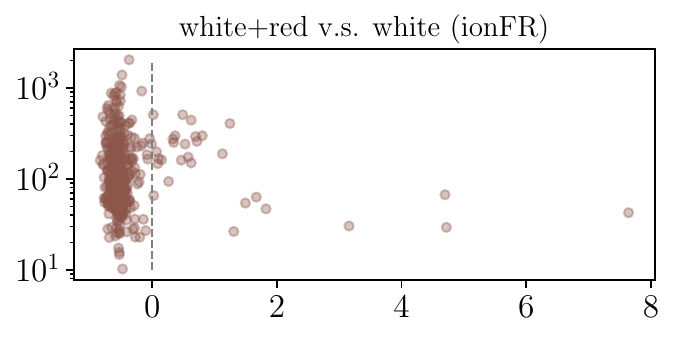}
    \caption{The scatter plots of the logarithm of Bayes Factor v.s. median of SNR for comparing different noise modellings across the full pulsar population that
passed initial selection. In the top row we compare the two ionosphere models ($415$ pulsars). 
    In the following rows, the results obtained with \texttt{spinifex} are presented in the left column ($513$ pulsars), while those with \texttt{ionFR} are shown in the right column ($415$ pulsars). The vertical dashed lines denote for $\ln\mathrm{BF} = 0$.}
    \label{fig: lnBF vs SNR}
\end{figure*}

In this subsection we provide comparison between different noise modellings by computing the logarithm of Bayes Factor through the Product Space method~\eqref{eq: likelihood hyper-parameter}. 
In Fig.~\ref{fig: lnBF vs SNR}, we show the scatter plot of the logarithm of Bayes Factors versus the median of SNR for comparing different models across the full pulsar population that passed initial selection.
Table.~\ref{tab:noise_compare} displays the corresponding Bayes factors for the ten pulsars with the highest SNR.


First of all, we observe that the Bayes factors obtained from comparing the two ionosphere models, \texttt{spinifex} and \texttt{ionFR}, are substantially greater than $1$ in most cases. This indicates that the ionospheric subtraction with \texttt{spinifex} generally yields a significantly better fit of the noise model to the data. The result is consistent with subsequent observations on the data properties.
When investigating the periodicity of the PA and RM time-series, we see clear evidence for seasonal variation in a majority of our pulsar's PA data which appears as a strong peak at a frequency around 31 nHz (1 year period). An example of this is shown in Fig.~\ref{fig:ion_comparison}. In general, we see that the default ionospheric modelling of \texttt{spinifex} is more effective at removing the seasonal variation in our data, with almost all pulsars (and especially those with high SNR that are used in further analysis) seeing a significant reduction in the power of the 1-year peak. We also note that the example shown in Fig.~\ref{fig:ion_comparison} is one of the extreme cases that highlights a major difference between the two models, and in general the majority of pulsars show similar behaviour. 

Second, we compare the IISM modellings to different orders ($2$nd to $4$th rows in Fig.~\ref{fig: lnBF vs SNR}). 
We first examine the effects of polynomial terms. For both ionospheric models, the corrected data for a considerable fraction of pulsars show significant evidence of a linear temporal trend, whereas the evidence for a quadratic trend is weaker. Next,
we examine the effect of including a harmonic term~\eqref{eqn:det_annual} with period of $1$ year in IISM modelling. For most pulsars, the data show no preference for the inclusion of such a term. 
This is particularly true when ionospheric effects are subtracted using \texttt{spinifex}, where the seasonal variation has already been accounted for (as discussed in the last paragraph). In contrast, in the comparison based on the \texttt{ionFR} model, some pulsars, such as J1935$+$1616, show clear evidence for a periodic component with a 1-year period in the Lomb-Scargle Periodogram even after subtraction of the ionospheric component (see Fig.~\ref{fig:ion_comparison}), which can be mainly attributed 
to inaccuracy of the ionospheric model at each observing epoch.

At last, we compare different random noise modellings. In the $5$th row of Fig.~\ref{fig: lnBF vs SNR},
we compare the full white‑noise model in Eq.~\eqref{eq:fullwn} with a minimal white‑noise model that includes only ionosphere‑corrected observational errors (i.e., with $\mathrm{EFAC}=1,\,\mathrm{EQUAD}=0$ in Eq.~\eqref{eq:fullwn}). 
The comparison reveals a clear division: more than half of the pulsars prefer the full noise model, whereas a smaller subset favours the minimal model. Pulsars in the latter group predominantly have $\mathrm{SNR}\lesssim 200$.
Then we compare the models with and without the red noise in the random noise modelling in the last row of Fig.~\ref{fig: lnBF vs SNR}. We find that, for most pulsars, the data does not show a significant preference for either model, and the Bayes factors are slightly below $1$ in many cases. Only 18 pulsars in the population ($<4\%$) have $\ln\mathrm{BF}\,_{\rm w}^{{\rm w}+{\rm r}} > 2.3$ (for \texttt{spinifex} model), which is a typical cut-off for estimating statistical significance in model comparisons.

Given that the data shows a strong preference on the ionospheric correction with the \texttt{spinifex} model (high Bayes Factors) in our noise model comparison, in the remainder of the paper we focus on the results found using the \texttt{spinifex} model. 

\begin{figure}
    \centering
    \includegraphics[width=0.95\columnwidth]{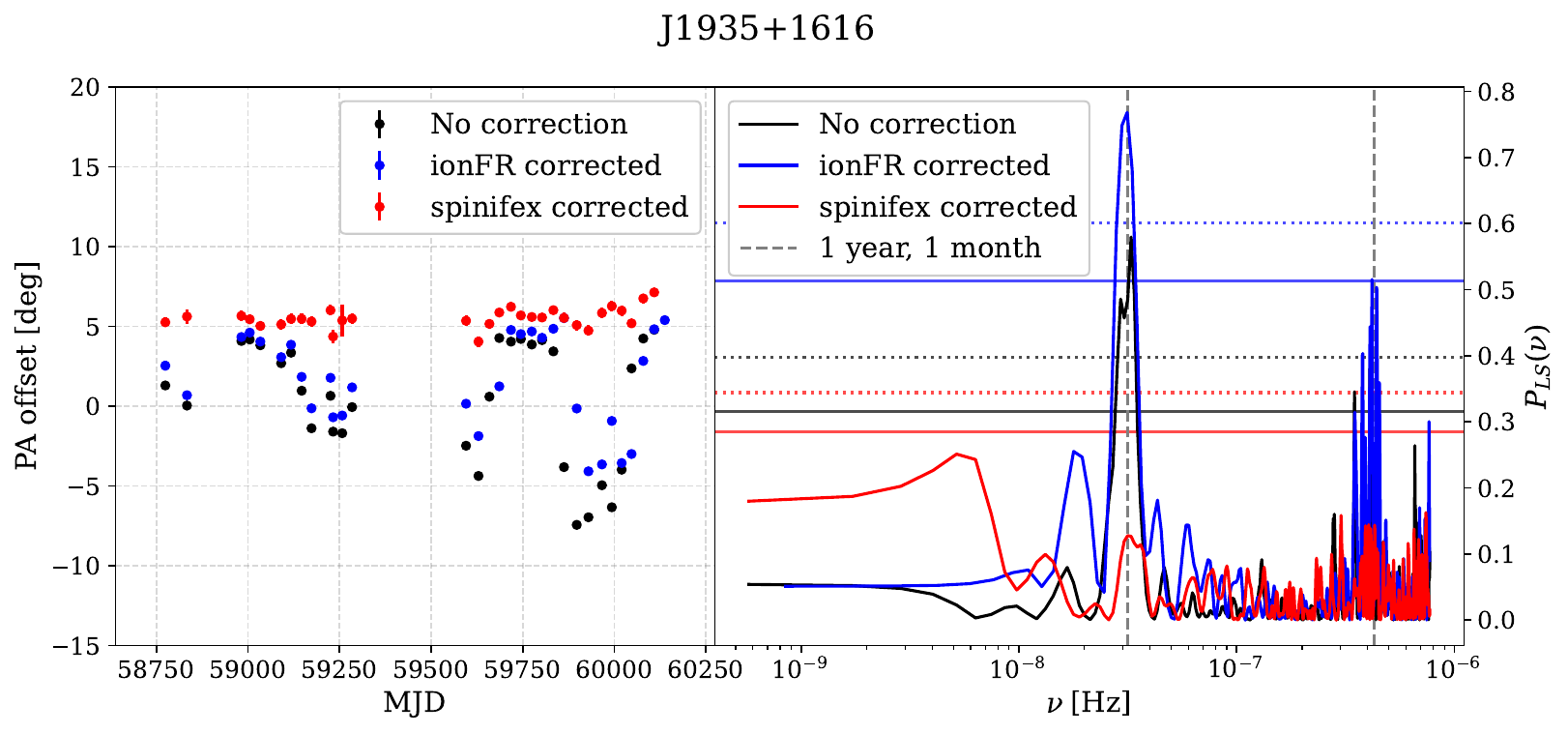}
    \caption{Accounting for ionospheric FR: two models \texttt{ionFR} (blue) and \texttt{spinifex} (red), for an example case of pulsar J1935+1616. The left panel shows the PA offset data for each observing epoch as calculated in Section~\ref{sec:data}, and the right hand panel shows the normalised power from the Lomb-Scargle Periodogram of each dataset. Vertical dashed lines show periods of interest related to the ionospheric variations, of a yearly seasonal period and an almost monthly ($\sim 27$ d) period associated with the solar rotation cycle. The horizontal solid and dotted lines of corresponding model colors show the corresponding peak height needed to reach the False Alarm Probabilities (FAPs) of 10\% and 1\%, respectively. }
    \label{fig:ion_comparison}
\end{figure}

\subsection{Noise parameter estimates}\label{sec:results-noise}

A set of MCMC sampling chains using a Bayesian likelihood model with the PA offset data and free parameters ${\mathrm{EFAC}, \mathrm{EQUAD}, S^{\rm r}, \Gamma}$ have been found for the full pulsar population. We find a mixed set of results, with a large majority of pulsars exhibiting white noise while being largely stable over the full observing window. Fig.~\ref{fig:SPA_summary} provides a basic set of summary statistics for these parameter estimates, and Table~\ref{tab:noise_analysis} provides the details of this procedure for a sample of pulsars (a full set of results can be found in the online supplementary material). Generally, the $\log_{10}(\mathrm{EFAC})$ parameter is constrained around 0 (with a mean of $-0.014$ for the population), which is the expected value in the case that the erorrbars on the PA offset are well-characterised. Upon visual inspection of the results, the samples that have poor constraints or very extreme values correlate with high scatter in the data or extremely large error bars. In terms of the red noise parameters, we see the majority of pulsars with either unconstrained parameters, or low estimated amplitudes. 

\begin{figure}[htb]
    \centering
    \includegraphics[width=0.95\columnwidth]{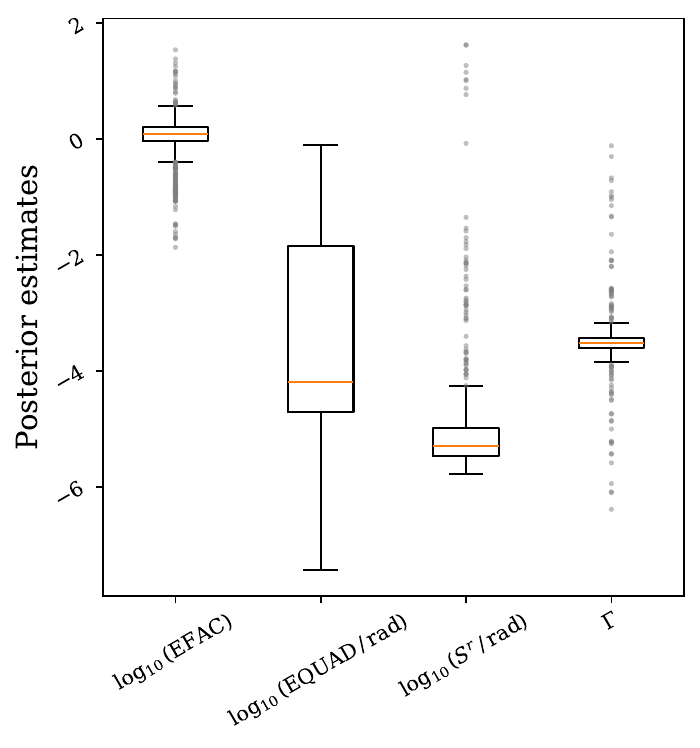}
    \caption{Summary of noise parameter estimation for the full population. For each parameter the sample consists of the median of the posterior probability distribution chain, computed via the PTMCMCSampler. Each box represents the standard interquartile range, median values are given by the orange horizontal line, whiskers are set at the usual 1.5 times the interquartile range, and outliers to this definition are plotted as individual dots.}
    \label{fig:SPA_summary}
\end{figure}

\begin{table*}[ht]
    \begin{tabular}{c|c c c c c c c }
        \multirow{2}{*}{}
        Pulsar name & $T_{\rm obs}$ [yr] & $N_{\mathrm{obs}}$  & med(SNR) & $\log_{10}(\mathrm{EFAC})$ & $\log_{10}(\mathrm{EQUAD/deg})$ & $\log_{10}(S^{\rm r}/\mathrm{deg})$ & $\Gamma$\\
        \hline
            J1935$+$1616 & 5.52 & 54 & 2034.32 & $-0.31_{-1.16}^{+1.13}$ & $-2.00_{-0.05}^{+0.05}$ & $-5.00_{-2.01}^{+1.52}$ & $-3.93_{-2.71}^{+2.58}$ \\	
            J1645$-$0317 & 5.46 & 73 & 1386.34 & $-0.72_{-0.87}^{+0.92}$ & $-1.63_{-0.06}^{+0.04}$ & $-4.06_{-2.68}^{+1.84}$ & $-2.58_{-3.22}^{+1.75}$ \\	
            J1913$-$0440 & 5.47 & 78 & 1065.82 & $-0.92_{-0.74}^{+0.78}$ & $-1.85_{-0.04}^{+0.04}$ & $-5.69_{-1.54}^{+1.75}$ & $-3.42_{-2.88}^{+2.40}$  \\	
            J1600$-$5044 & 5.46 & 75 & 1023.81 & $0.31_{-1.51}^{+0.81}$  & $-1.55_{-2.05}^{+0.06}$ & $-5.48_{-1.69}^{+1.81}$ & $-3.47_{-2.88}^{+2.40}$  \\	
            J1932$+$1059 & 5.52 & 75 & 922.00  & $-0.01_{-0.04}^{+0.04}$ & $-5.03_{-1.98}^{+2.07}$ & $-5.64_{-1.57}^{+1.72}$ & $-3.55_{-2.87}^{+2.41}$ \\	
            J1709$-$4429 & 6.11 & 160 & 889.92 & $1.31_{-0.02}^{+0.03}$  & $-5.28_{-1.84}^{+1.92}$ & $-3.98_{-1.99}^{+0.97}$ & $-4.36_{-2.42}^{+2.21}$ \\	
            J1820$-$0427 & 5.46 & 76 & 875.35  & $0.95_{-1.62}^{+0.07}$  & $-2.31_{-3.85}^{+0.65}$ & $-5.70_{-1.57}^{+1.69}$ & $-3.35_{-2.92}^{+2.32}$ \\	
            J1752$-$2806 & 5.48 & 73 & 862.29  & $1.16_{-1.97}^{+0.44}$  & $-1.22_{-0.44}^{+0.07}$ & $-4.40_{-2.45}^{+1.56}$ & $-4.09_{-2.68}^{+2.69}$ \\	
            J1359$-$6038 & 6.00 & 76 & 838.29  & $-0.49_{-1.03}^{+1.03}$ & $-2.22_{-0.07}^{+0.05}$ & $-2.87_{-0.87}^{+0.35}$ & $-3.72_{-2.60}^{+1.50}$ \\	
            J0820$-$1350 & 5.94 & 48 & 815.81  & $1.09_{-0.05}^{+0.05}$  & $-4.75_{-2.23}^{+2.17}$ & $-5.47_{-1.68}^{+1.76}$ & $-3.62_{-2.84}^{+2.47}$  \\	
        $\vdots$ & & & & & & & \\
        \hline
    \end{tabular}
    \caption{Summary of MCMC parameter sampling and estimation results for a sample of pulsars. Columns show, from left to right: Pulsar name, total observation timespan (i.e. MJD(last observation) - MJD(first observation)), number of individual time series data points, median SNR of all data points, parameter estimates for \{EFAC, EQUAD, $S^{\rm r}$, $\Gamma$\}. The full table can be requested from the authors.}
    \label{tab:noise_analysis}
\end{table*}

In Fig.~\ref{fig:J1853-0004_spa_corner} we show an instructive example output of the posterior probability densities from the MCMC sampling, carried out with the \texttt{PTMCMCSampler}~\cite{justin_ellis_2017_1037579}. In this particular example we see clear evidence of red noise, given by the amplitude parameter $\log_{10}(S^{\rm r}) = -2.75^{+0.44}_{-0.35}$ and index $\Gamma = -6.39\pm^{1.45}_{1.09}$. When comparing this result visually to the data shown in Fig.~\ref{fig:J1853-0004_spinifex_vs_observed}, we see a clear upturn in PA offset midway through the observations. Corresponding RM estimates for this pulsar are provided in Fig.~\ref{fig:J1853-0004_RMVariations_LSP}, and we note a previously measured linear slope in RM synth data of $1.54 \pm 0.31\; \mathrm{rad \, m^{-2} \, year^{-1}}$~\cite{Keith:2024kpo}, though the domain of that study only includes observations up until MJD $\sim 60000$, and the recent data suggests the linear trend is no longer suitable for this pulsar. 
Further, a computed Bayes Factor of $\ln(\mathrm{BF}_{\rm w}^{{\rm w}+{\rm r}}) = 5.81$ shows a clear preference for the inclusion of the red noise model to the data. Finally, we see in this pulsar another case of good modelling for the yearly seasonal effects of the ionosphere, given by the total reduction in the peak of the LSP at the 1-year frequency when the \texttt{spinifex} ionospheric model is subtracted from the data, though there remains a small excess of power on a monthly timescale, possibly associated with the solar rotation cycle. For comparison, we also show the same analysis for the pulsar J1949$-$2524 in Fig.~\ref{fig:J1949-2524_RMVariations_LSP}, which shows no evidence for red noise (with $\ln(\mathrm{BF}_{\rm w}^{{\rm w}+{\rm r}}) = -0.58$ and unconstrained amplitude $S^{\rm r}$). This is again quite clear visually in the data, which appears to be stable over the entire observation timespan, and this behaviour is the most typical in the population. 

\begin{figure}
    \centering
    \includegraphics[width=0.95\columnwidth]{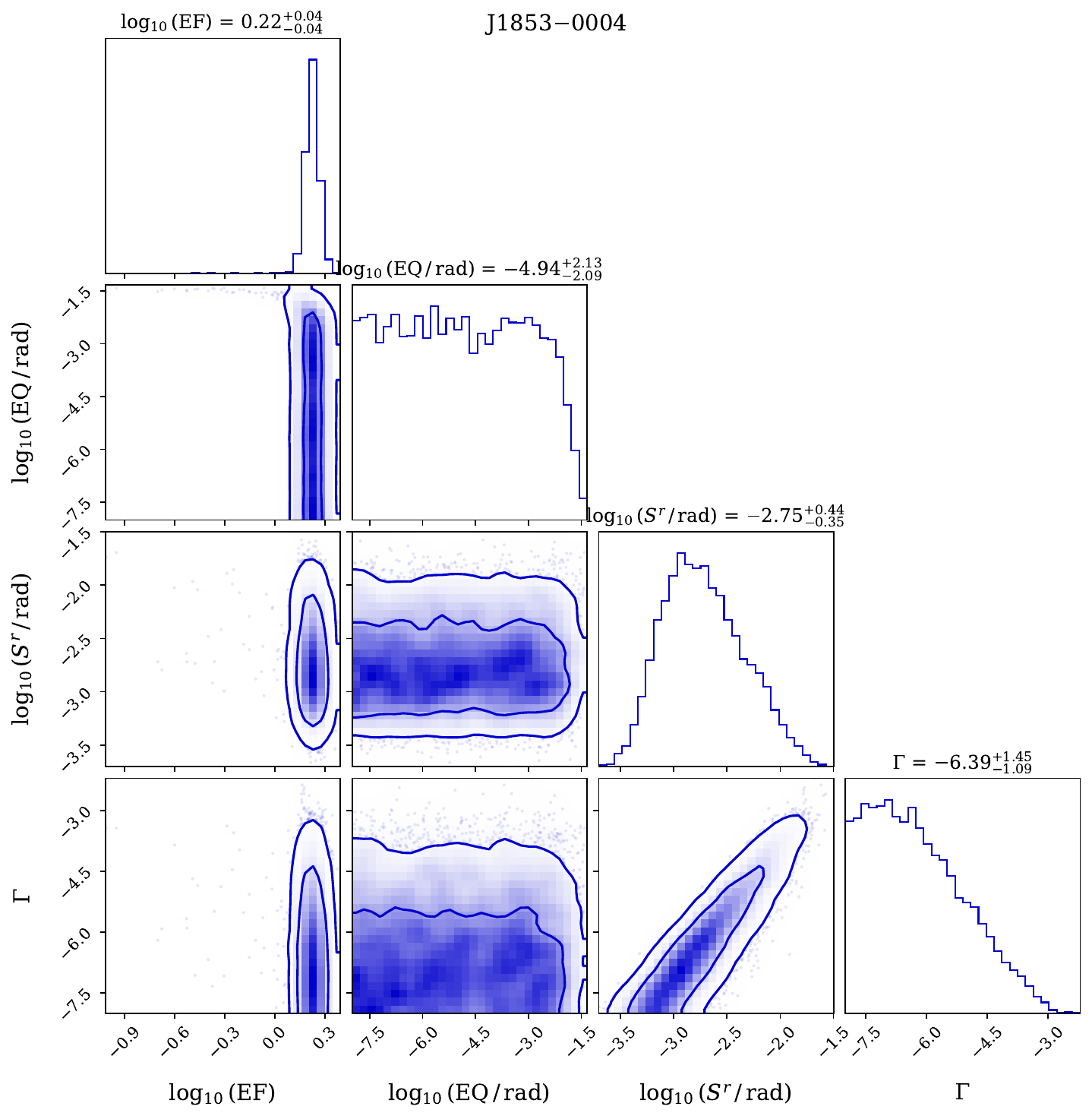}
    \caption{Bayesian parameter sampling for white and red noise models with the marginalised likelihood in Eq.~\eqref{eq: lnLm}, parameterised by EFAC, EQUAD and $S^{\rm r}$, $\Gamma$ respectively. This is an example output for the pulsar J1853$-$0004, which shows strong evidence of red noise.}
    \label{fig:J1853-0004_spa_corner}
\end{figure}

\begin{figure}
    \centering
    \includegraphics[width=0.95\columnwidth]{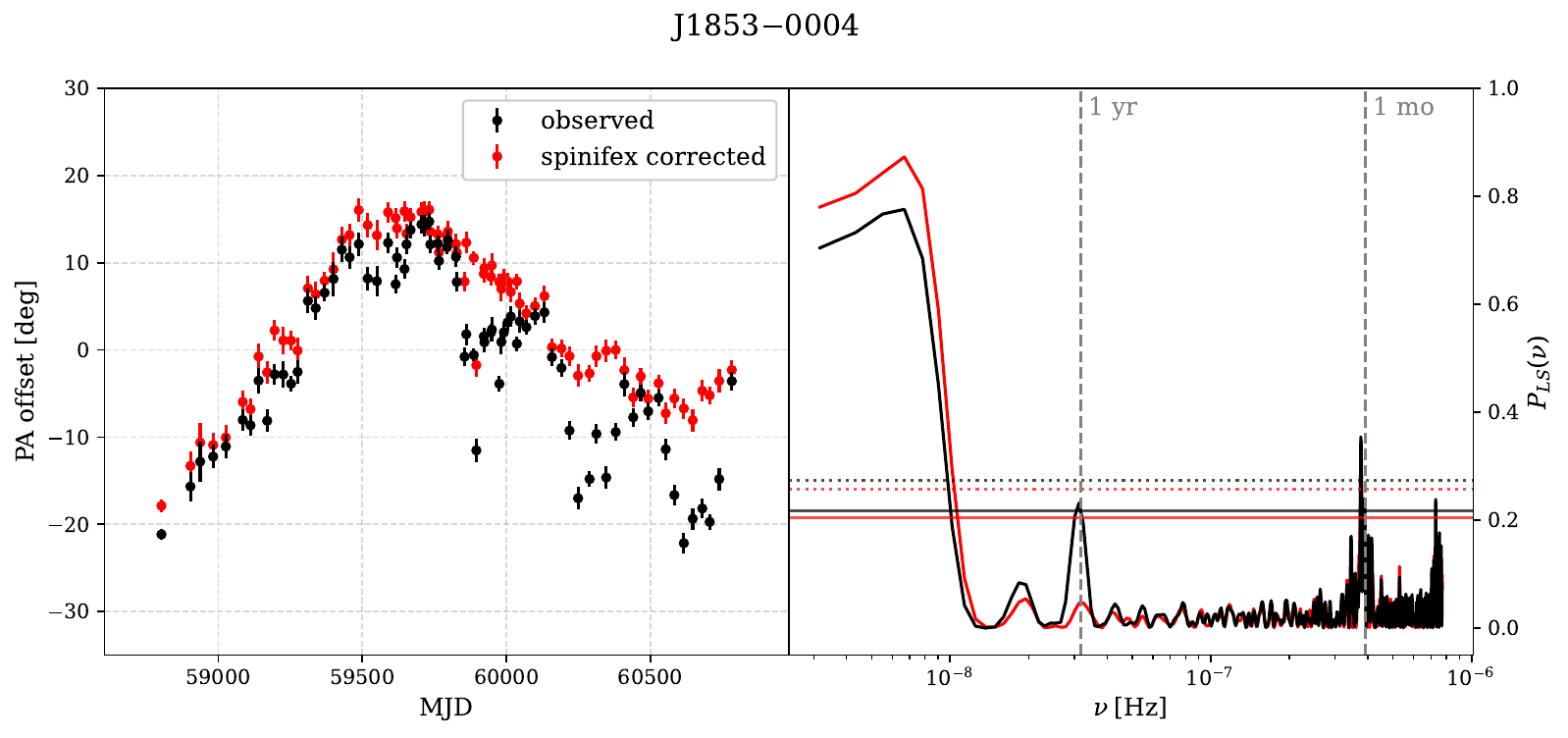}
    \caption{PA offset behaviour for pulsar J1853$-$0004. Left Panel: PA offset time series data without ionospheric correction (black) and with \texttt{spinifex} model subtraction (red). Right panel: LSP of the data from the left panel, with respective color scheme. Also shown are 1-year and 1-month periods in vertical dashed lines, and the LSP FAPs of 10\% and 1\% by the horizontal solid and dotted lines, respectively.}
    \label{fig:J1853-0004_spinifex_vs_observed}
\end{figure}

\begin{figure}
    \centering
    \includegraphics[width=0.95\columnwidth]{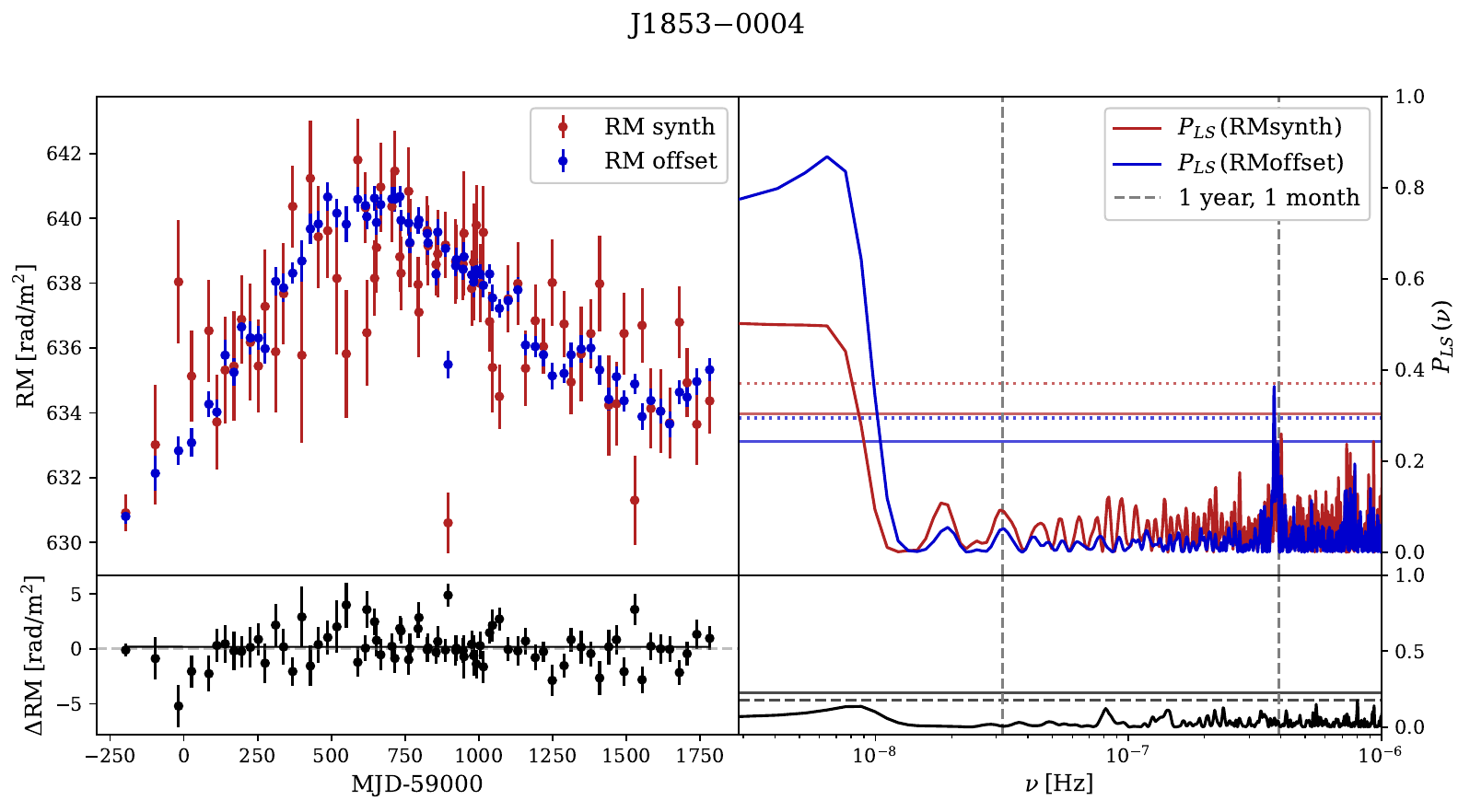}
    \caption{Variations in the RM values obtained through `RMSynth' (red) and `RM offset' (blue) techniques (see Section~\ref{sec:RM} for details). The left-hand panels show the RM values themselves with associated uncertainty bars, and the right-hand panels show the computed Lomb-Scargle periodograms from each dataset. Vertical dashed lines show the frequencies corresponding to yearly and monthly periods, and the horizontal solid and dotted lines of each colour show the computed FAP of 10\% and 1\%, respectively. The bottom panels show the residual data points, i.e. RM offset - RM synth. }
    \label{fig:J1853-0004_RMVariations_LSP}
\end{figure}

\begin{figure}
    \centering
    \includegraphics[width=0.95\columnwidth]{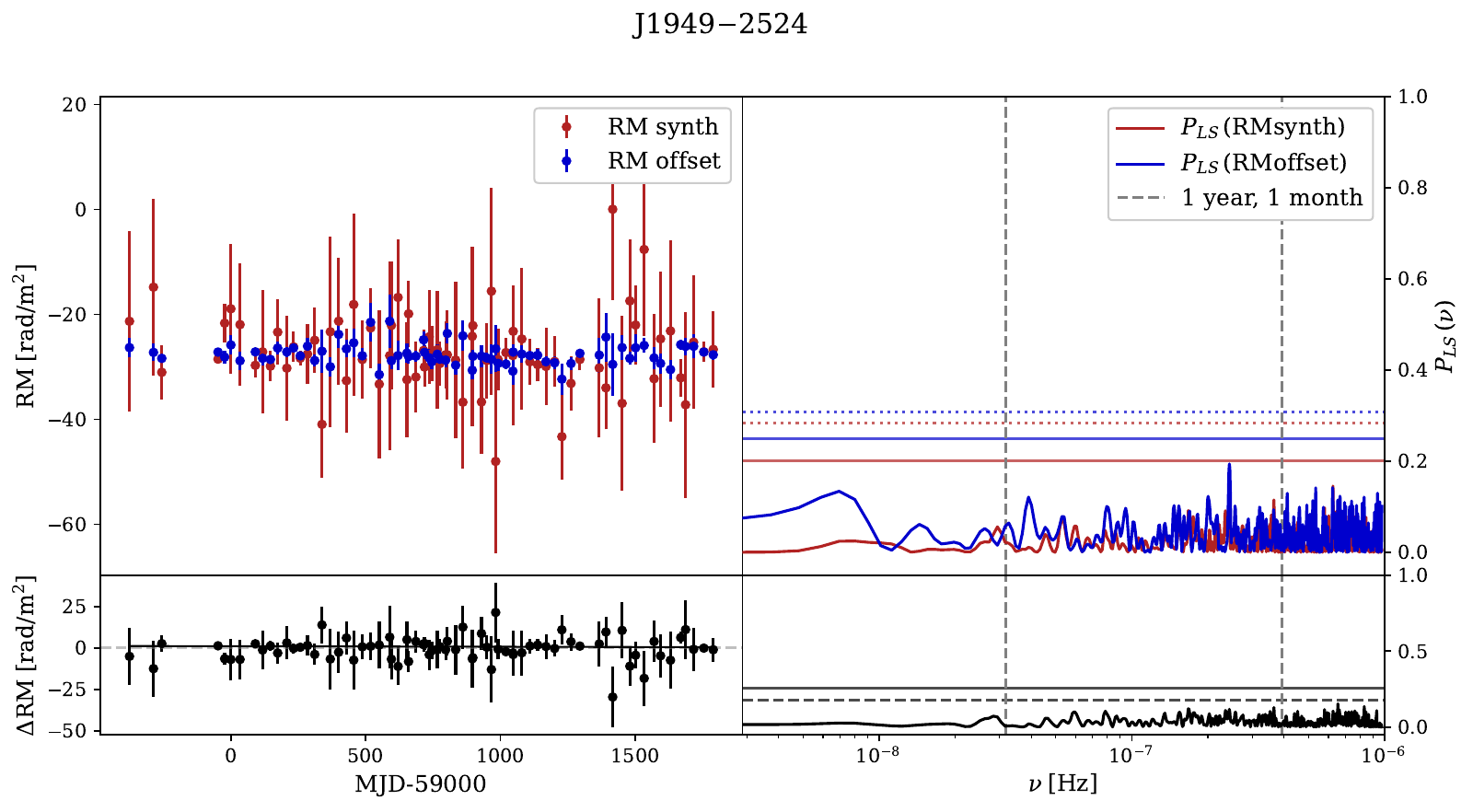}
    \caption{Same as Fig.~\ref{fig:J1853-0004_RMVariations_LSP}, but for pulsar J1949$-$2524. The stable behaviour of $\Delta \mathrm{RM}$ exhibited here is observed most typically in the population.}
    \label{fig:J1949-2524_RMVariations_LSP}
\end{figure}

\subsection{RM variations}\label{sec:results-RM}

To analyse the linear trends in the RM residuals, we have used the Levenberg-Marquardt algorithm from the lmfit Python package. This algorithm also provides us with a set of fit statistics to measure how successful the fit is for each pulsar, including standard $\chi^2$ statistics and estimates on parameter covariance. After analysing the trends in the RM residuals, we find that only 27 pulsars (5\% of total dataset) have significant evidence of a non-zero linear slope, chosen to be above the threshold of $3\sigma$ away from 0. Of these 27, most have reduced $\chi^2$ values $\approx 1$ for the linear fit, with only 3 having a value of $\vert\chi^2_{\rm r}\vert > 10$, which indicates the linear function as unsuitable to fit the data. Upon inspection of these pulsars, it seems likely that a combination of the small errors on these data points and the relatively large scatter leads to these large $\chi^2_{\rm r}$ values, which could indicate that the error estimates have been underestimated due to intrinsic PA variability of the pulsar itself because of jitter or otherwise. We note~\cite{Keith:2024kpo} found that $\sim 14\%$ of a very similar pulsar population, with largely the same TPA data, had significant evidence of a linear slope in the standard RM synth measurements. 

As discussed in~\cite{wangPropagationinducedFrequencydependentPolarization2025}, a more general model of radiative transfer that employs the Faraday mixing scenario, i.e. a combination of Faraday rotation, conversion and absorption effects, might be necessary in these pulsars. These effects are significant for propagation within areas of strong magnetic fields or dense plasmas, and a more detailed study of the spectro-polarimetric properties of this dataset might yield further insights into whether our observed RM variations are dominated by these or other effects. 

One clear result from this analysis is that we find no strong evidence for a periodicity in the RM residuals. This implies that there is no common characteristic wavelength-independent oscillation of the PA, which could be a signal of an intervening ALDM field. In many pulsars we do see excess power in the LSP at yearly and monthly periods, however there are viable astrophysical explanations for these cases in the form of ionospheric density fluctuations caused by solar cycles. We note that although excess power at monthly periods does correspond to our average observation cadence, the LSP still provides a robust result on this timescale because of the non-uniformity of the timeseries data~\cite{vanderplasUnderstandingLombScarglePeriodogram2018}, and that there is a known physical mechanism for this result based on the Sun's rotational period of 27 days.

\subsection{Pulsars of Interest}\label{subsec: puslars of interest}

In this section we present a small set of pulsars that exhibit unique or non-standard behaviour in their time-series, and which may warrant further investigation. 

\begin{description}
    \item[J1428$-$5530 (Fig.~\ref{fig:J1428-5530_PAmodes})] The PA offset data of this pulsar (also known as B1424$-$55) shows clear evidence of two orthogonal polarisation modes (OPM), seen in Fig.~\ref{fig:J1428-5530_PAmodes}. This behaviour has been observed in pulsar data before, for example in Ref.~\cite{oswaldPulsarPolarizationBroadband2023}, and a recent study has attempted to model the observations using a combination of coherent and incoherent combination of the two polarisation modes~\cite{oswaldPulsarPolarizationPartialcoherence2023}.In the relatively short observations different OPM dominate, which are offset by 90 degree in PA.  We note that Ref.~\cite{oswaldPulsarPolarizationBroadband2023} acknowledges the orthogonal jump in this pulsar, with further characterisation of non-zero circular polarisation and frequency evolution of the intensity profile, PA and circular polarisation profile shape, and fraction of linear polarisation. Other than the OPMs, there is no clear evidence for variability in the PA offset or RM offset data.\\
    
    \begin{figure}
        \centering
        \includegraphics[width=0.95\columnwidth]{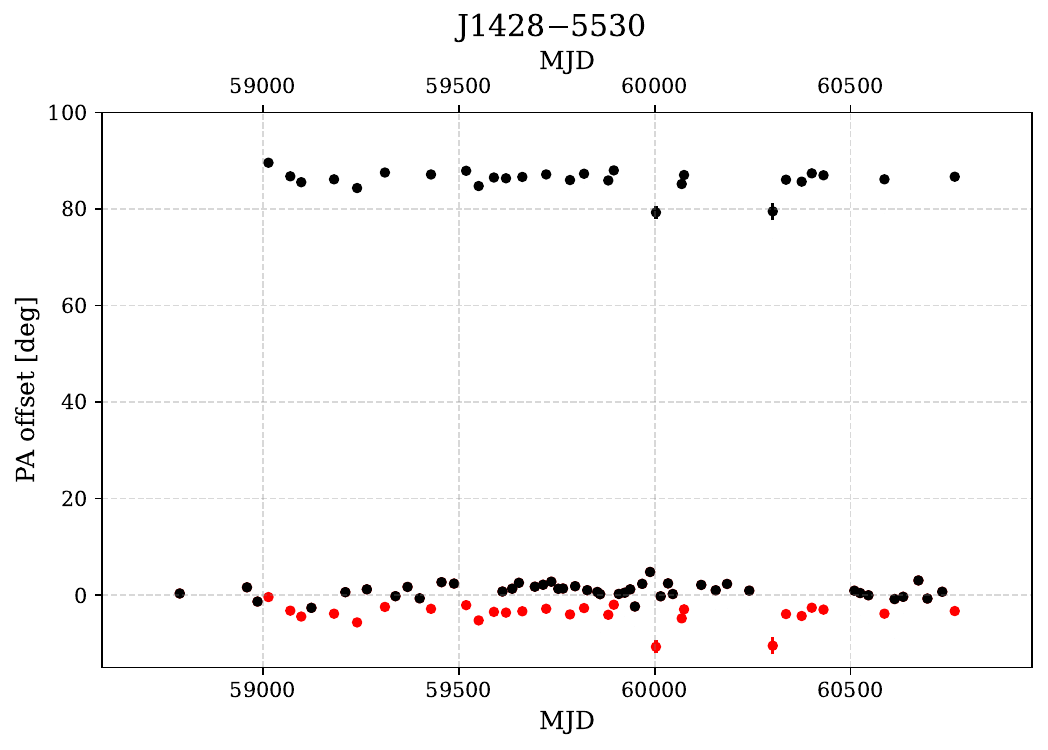}
        \caption{PA offset data for pulsar J1428$-$5530. Two orthogonal polarisation modes are clearly visible, with original data represented by black data points and the 90 degree ``jumps'' represented by the red data points.}
        \label{fig:J1428-5530_PAmodes}
    \end{figure}


    \item[J1141$-$6545 (Fig.~\ref{fig:J1141-6545_RMvariations})] In this pulsar we note that the computed RM synth values align well with the values reported in Ref.~\cite{oswaldThousandPulsarArrayProgrammeMeerKAT2025}. However, there is a visually clear slope in the difference of RM synth and RM offset, which when fitted with a linear trend has a fitted slope parameter of $a = -0.518 \pm 0.040 \; \mathrm{rad \, m^{-2} \, year^{-1}}$, which is non-zero to $\approx 12\sigma$. The reduced $\chi^2$ statistic for this fit is 1.6, indicating a suitable linear fit to this data. We also note that, because the measured RM synth data appears stable in time, there is no measured RM slope reported by~\cite{Keith:2024kpo}. According to our assertions in Section~\ref{sec:RM}, variability in RM offset values that does not also appear in RM synth could be interpreted as a changing wavelength-independent contribution to rotation of the PA. An alternative explanation for this trend could be a gradual changing of the intrinsic PA angle of the pulsar, which would not affect usual RM measurements. 
    Binary effects could play a role, since the pulsar is has a white dwarf companion\cite{bbv08}.
    We leave further interpretations of this effect to future work.  \\
    
    \item[J1105$-$6107 (Fig.~\ref{fig:J1105-6107_RMvariations})] This pulsar has similar visual behaviour to J1141$-$6545. In this case we find a slope to the residual RM of $a = 0.654 \pm 0.066\; \mathrm{rad \, m^{-2} \, year^{-1}}$, with a significance of $\approx 10\sigma$, though the goodness-of-fit estimate of $\chi_{\rm r}^2 = 4.2$ shows a weaker fit to a linear function. Of interest here is that also the total intensity pulse profile shape of this pulsar is known to gradually evolve over time\cite{bkj+16}.\\
    
    \item[J1048$-$5832 (Fig.~\ref{fig:J1048-5832_RMvariations})] In this pulsar (which is also known as B1046$-$58) we also see a deviation between the trends of RM synth and RM offset, though unlike J1141-6545 and J1105-6107, in this case the PA offset values seem stable over the time-series while the RM synth trend is very clearly linear with a positive slope. Here we find a slope to the residual RM of $a = -0.835 \pm 0.047\; \mathrm{rad \, m^{-2} \, year^{-1}}$, significant to almost $18\sigma$ and consistent with \cite{Keith:2024kpo}.\\
    \item[Long-term variability] The following set of pulsars show evidence of long-term variability in both RM measures (i.e. both RM synth and RM offset): J1056$-$6258, J1114$-$6100, J1825$-$1446, J1915+1009, and J1359$-$6038, which is consistent with our MCMC results. For \texttt{spinifex} model, the $\ln\,\mathrm{BF}$ comparing IISM modellings with linear function of time ($1$st-order polynomial) to that as a constant ($0$th-order polynomial) take the values $0.45$, $8.08$, $8.21$, $6.57$, and $5.51$, respectively.
    Apart from J1359$-$6038, these pulsars all have statistically significant RM slopes measured in~\cite{Keith:2024kpo}. Since these temporal trends are present in both RM measures, which suggests a likely IISM-related cause, more sophisticated physical modelling may be necessary in future work to understand the precise nature of this variability.   
\end{description}

\begin{figure*}
    \centering
    \hfill
    \begin{subfigure}[b]{0.475\textwidth}
        \includegraphics[width=\textwidth]{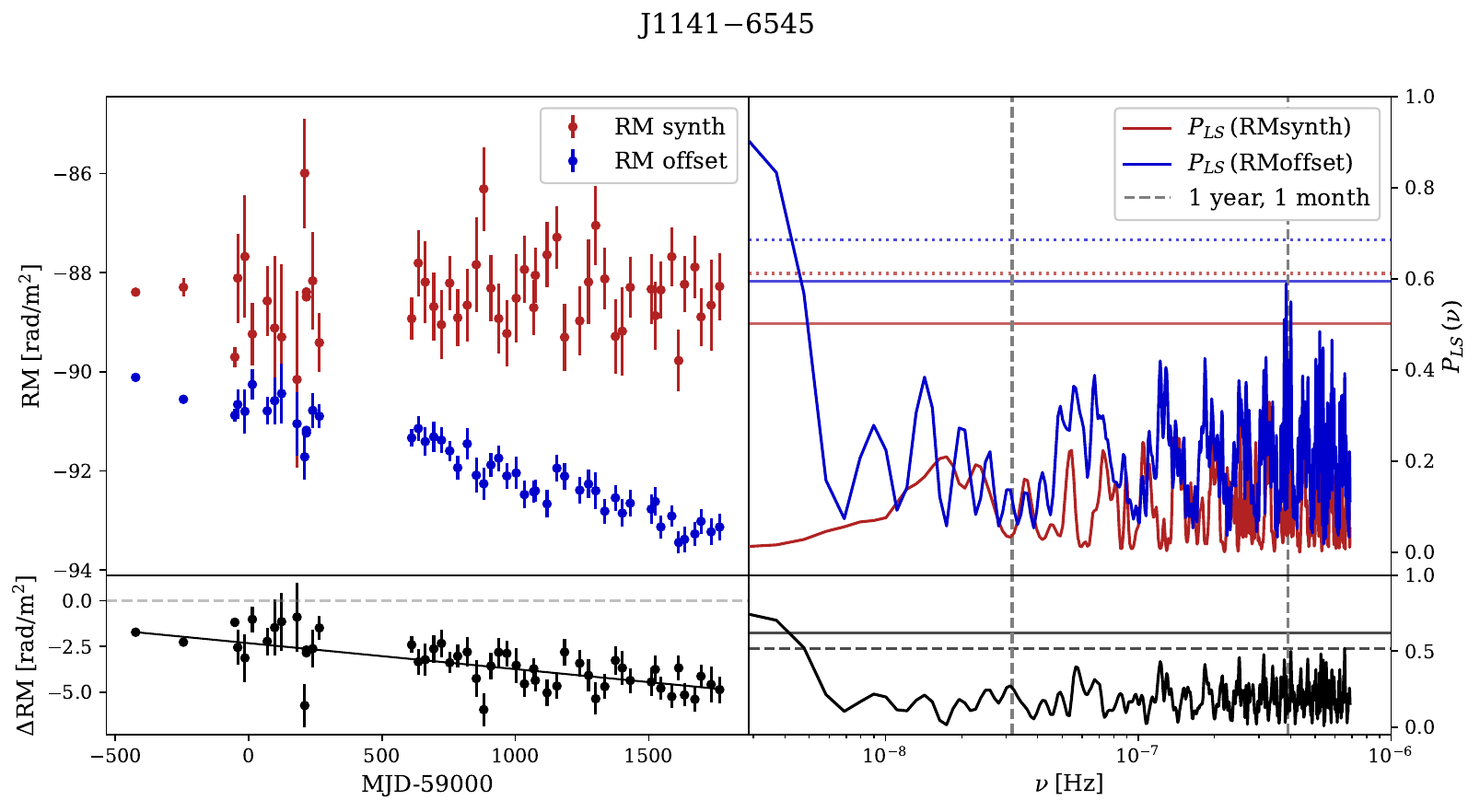}
        \caption{J1141$-$6545}
        \label{fig:J1141-6545_RMvariations}
    \end{subfigure}
    \begin{subfigure}[b]{0.475\textwidth}
        \includegraphics[width=\textwidth]{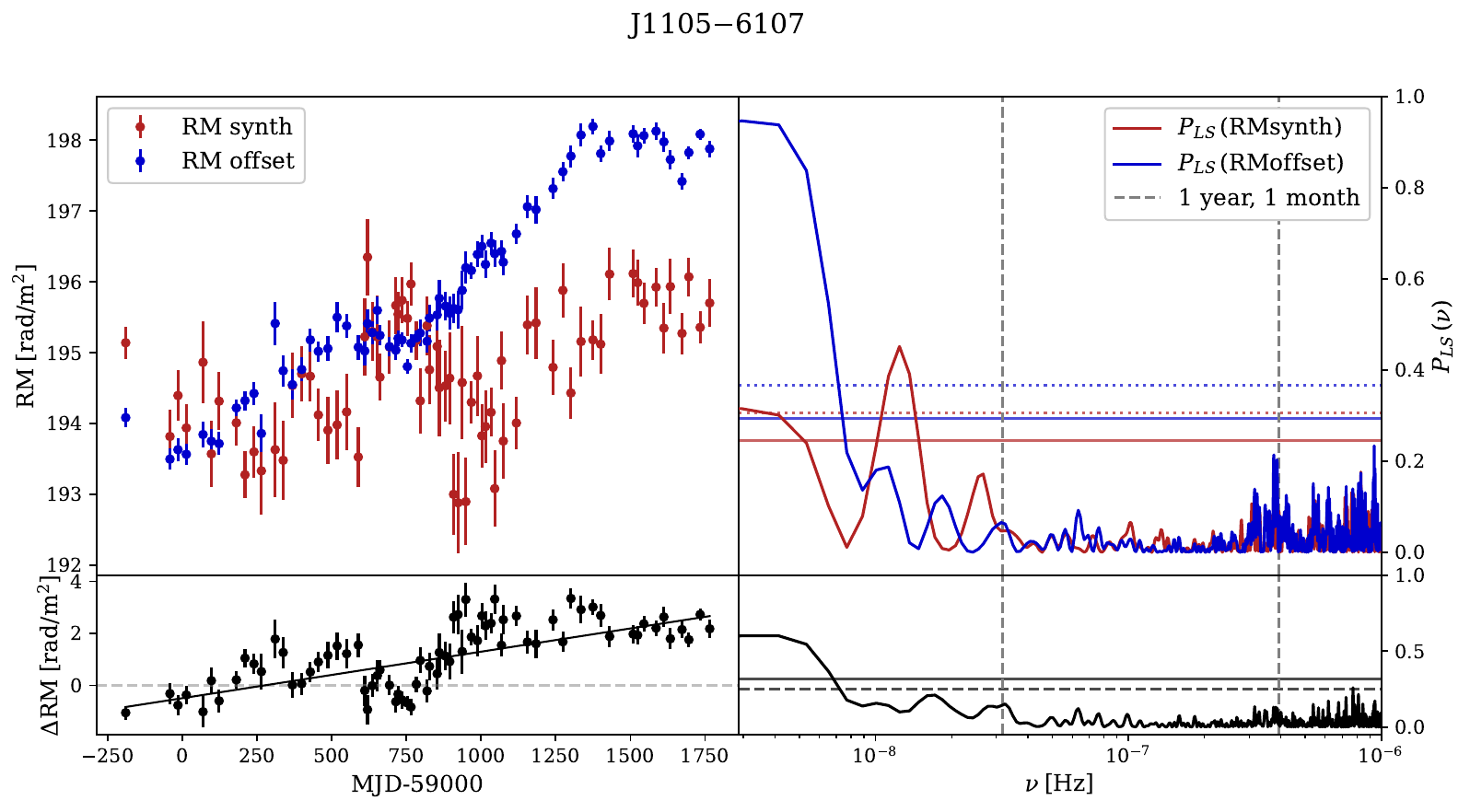}
        \caption{J1105$-$6107}
        \label{fig:J1105-6107_RMvariations}
    \end{subfigure}

    \vspace{0.2cm}
    \hfill
    \begin{subfigure}[b]{0.475\textwidth}
        \includegraphics[width=\textwidth]{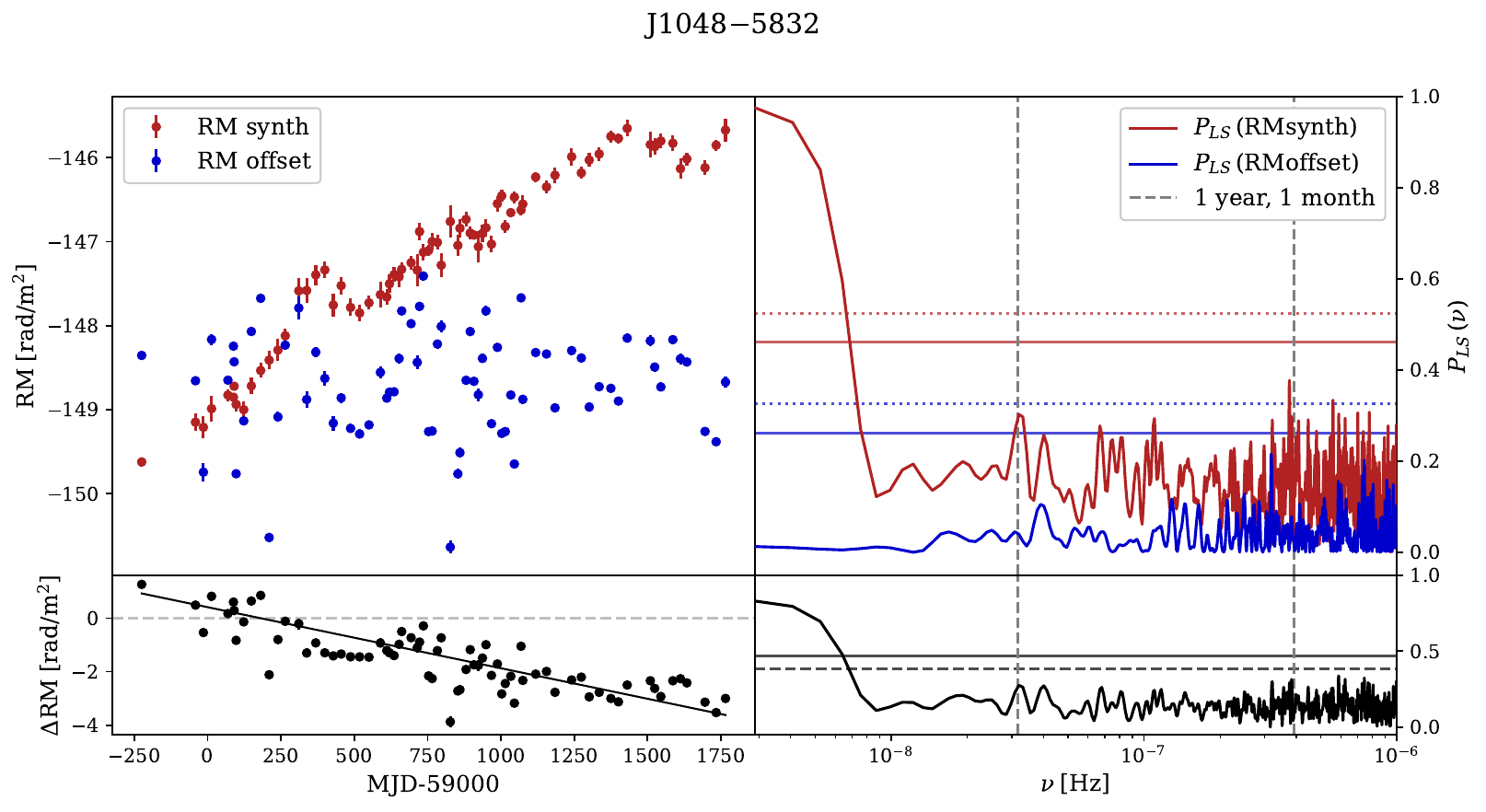}
        \caption{J1048$-$5832}
        \label{fig:J1048-5832_RMvariations}
    \end{subfigure}
    \begin{subfigure}[b]{0.475\textwidth}
        \includegraphics[width=\textwidth]{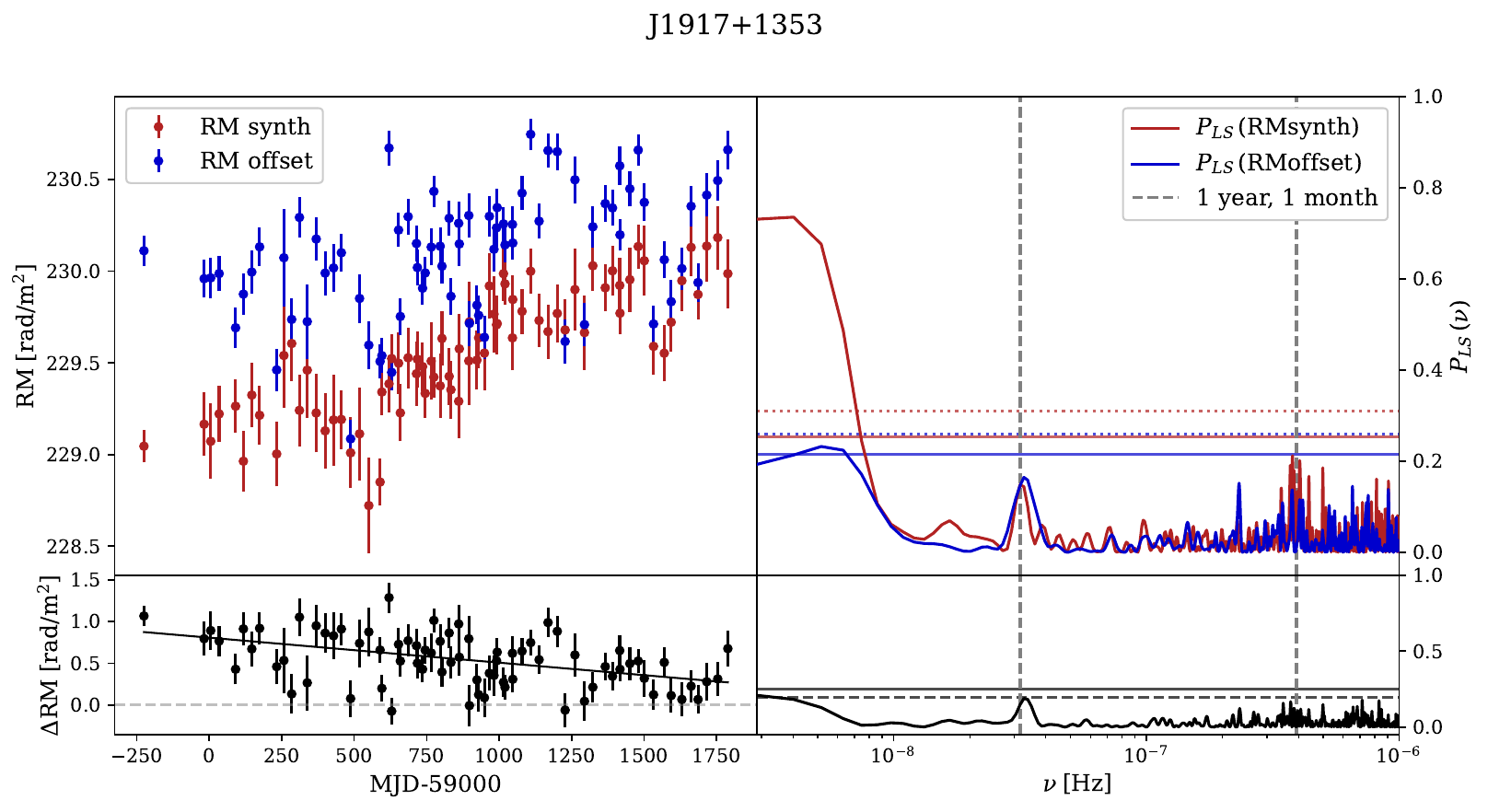}
        \caption{J1917+1353}
        \label{fig:J1917+1353_RMvariations}
    \end{subfigure}

\caption{Variations in RM obtained for `Pulsars of Interest'. Information shown in each panel is the same as Fig.~\ref{fig:J1853-0004_RMVariations_LSP}. }\label{fig: multi_RMVariations}
\end{figure*}

\section{Conclusion}

This work, which is part of a larger study into the use of the first regular-pulsar PPA, namely the MeerKAT TPPA, as a probe of ultralight ALDM, presents an analysis of the temporal behaviour of the data and general trends within a large population of 513 pulsars. We defined a parameter, which we called the PA offset, that tracks the variation in the PA compared to a median template PA value, per each observation epoch over a timespan of roughly 6.5 years. The analysis of this dataset followed a simple Bayesian likelihood approach, with two primary long-term variation effects: a second-order polynomial to track long-term fluctuations in IISM, and a harmonic function with a one-year period to track orbital changes in line-of-sight, and two generic noise terms: white noise parameterised by EFAC and EQUAD, and red noise parameterised by an amplitude and spectral index. We identified several trends in the results of the MCMC parameter sampling, with the majority of pulsars in the sample only exhibiting white noise, and only 18 pulsars showing clear evidence of red noise. 

We also presented a comparison between two estimates of the RM, which represents the characteristic magnitude of PA rotation. As the derivations of these two estimates were completely independent and relied on different assumptions about the underlying wavelength-dependence of the rotation, their comparison provides a new probe into the potential contributions to the PA that are non-Faraday-like. We find no strong evidence for oscillation of the residual RM, 
and in the vast majority of our population we find a stable residual RM - indicating that the dominant component of RM is indeed FR. However, we find evidence of deviation from this scenario in some pulsars; specifically, we find two OPMs in the PA offset data of J1428$-$5530, and we find statistically significant evidence of a linear variation in the RM when determined via RM synthesis and via our frequency-independent templating technique in 3 pulsars, most significantly in J1141$-$6535, but also in J1917+1353, J1651$-$4246 and J1105$-$6107.   

We also found that the use of the \texttt{spinifex} ionospheric model is generally successful in removing seasonal variations of in the PA in this dataset, though shorter timescale variations on the order of one month (which may be associated with physical effects like the solar rotational cycle) persist in many pulsar's data. More sophisticated models and precise TEC maps may better account for the ionospheric FR components that may be missed with the simple SLM used in this work~\cite{poraykoValidationGlobalIonospheric2023}, and with more sensitive data, more detailed ionospheric analysis should be warranted.

\section{Acknowledgements}
We thank Roland Crocker, Geoff Beck, Chris Gordon and Steven Murray for helpful comments and suggestions on this work. Z.-Y. Yuwen acknowledges the support from the Program of China Scholarship Council Grant No. 202404910329, and is also supported by the Young Scientist Training (YST) Program at the APCTP through the Science and Technology Promotion Fund and Lottery Fund of the Korean Government. Y.-Z. Ma acknowledges the support from South Africa's National Research Foundation under Grants No.~150580, No.~CHN22111069370, No.~ERC250324306141. 
J.R. is supported in part by the National Natural Science Foundation of China under Grant No. 12435005.
X.X. is funded by the grant CNS2023-143767. 
Grant CNS2023-143767 funded by MICIU/AEI/10.13039/501100011033 and by European Union NextGenerationEU/PRTR. We acknowledge computational cluster resources at the Centre for High-Performance Computing, Cape Town, South Africa.
The MeerKAT telescope is operated by the South African Radio Astronomy Observatory, which is a facility of the National Research Foundation, an agency of the Department of Science and Innovation. MeerTime data are housed and processed on the OzSTAR supercomputer at Swinburne University of Technology. Pulsar research at Jodrell Bank Centre for Astrophysics and Jodrell Bank Observatory is supported by a consolidated grant from the UK Science and Technology Facilities Council (STFC). 

\appendix

\section{Marginalisation of IISM parameters}\label{sec:marginalisation}

In Section~\ref{sec:likelihoods}, we described the Bayesian likelihood used for analysing the noise models. 
In this Appendix, we provide a detailed derivation of the marginalizsation of the IISM parameters. To begin with, let us recall the full likelihood function given in Eq.~\eqref{eq: lnL},
\begin{align}\label{eq: lnL_appendix}
        \ln \mathcal{L} ~=& -\frac{1}{2}\left(\delta \mathrm{PA} - M\psi\right)^{\transp} C^{-1} \left(\delta \mathrm{PA} -M\psi\right) - \frac{1}{2}|2\pi C|~ \\
        =& -\frac{1}{2}(M\psi)^\transp C^{-1} (M\psi) - \frac{1}{2}\left(\delta \mathrm{PA}\right)^\transp C^{-1} \delta \mathrm{PA} \nonumber \\
        & + \left(\delta \mathrm{PA}\right)^\transp C^{-1} (M\psi) - \frac{1}{2}\ln |2\pi C|~,
\end{align}
For the full analysis, one can marginalize the parameters $\psi$ by integrating the full likelihood over the corresponding parameter space. Without any prior knowledge about $\psi$, one may assume that all points in the parameter space carry equal weight, which is analogous to adopting an uninformative prior, and in this case the likelihood reduces to a Gaussian-like function of $\psi$. It turns out the integration of $\psi$ can be carried out analytically. We first define $C_M\equiv M^\transp C^{-1} M$ and $\delta\mathrm{PA}^c \equiv M^\transp C^{-1}\delta \mathrm{PA}$, then Eq.~(\ref{eq: lnL_appendix}) can be re-written as
\begin{align}
\begin{aligned}\label{eq:lnL_intermediate}
    \ln \mathcal{L} =& -\frac{1}{2} \left(\psi - C_M^{-1} \delta \mathrm{PA}^c \right)^\transp C_M \left(\psi - C_M^{-1} \delta \mathrm{PA}^c \right)  \\
    & + \frac{1}{2} (\delta\mathrm{PA}^c)^\transp C_M^{-1} \delta\mathrm{PA}^c \\
    &-\frac{1}{2}\left(\delta \mathrm{PA}\right)^\transp C^{-1} \delta\mathrm{PA} - \frac{1}{2}\ln |2\pi C| .
\end{aligned}
\end{align}
We then integrate $\mathcal{L}_\mathrm{m}=\int\mathcal{L}\md\psi$ by utilizing the fact
\begin{eqnarray}
    &&\int \exp\left(-\frac{1}{2}\left(\psi - \psi' \right)^\transp C_M \left(\psi -\psi' \right) \right)\md \psi \nonumber \\
    =&& \sqrt{|2\pi C^{-1}_{M}|} \nonumber \\
    =&& \sqrt{|2\pi C_{M}|^{-1}}+{\rm const.} \label{eq:lnL_intermediate2} 
\end{eqnarray}

Using Eq.~(\ref{eq:lnL_intermediate2}), the marginalization becomes
\begin{align}\label{eq: lnLm_appendix}
    \ln \mathcal{L}_\mathrm{m} &=  \frac{1}{2}(\delta\mathrm{PA}^c)^\transp C_M^{-1} \delta\mathrm{PA}^c  - \frac{1}{2}\ln |2\pi C_M| \\
    &\quad~ - \frac{1}{2}\left(\delta \mathrm{PA}\right)^\transp C^{-1} \delta\mathrm{PA} - \frac{1}{2}\ln |2\pi C| + \mathrm{const.} \nonumber,
\end{align}
which we used in most cases throughout this analysis and in the ALDM search found in our accompanying work~\cite{Yuwen:2025}. 
A constant term in $\ln \mathcal{L}_\mathrm{m}$ does not affect the result and can be removed by performing a normalization, therefore the last constant term in Eq.\eqref{eq: lnLm_appendix} can be neglected. In total, for each individual pulsar there are $11$ free parameters included in the marginalized Bayesian model that uses the likelihood in Eq.~\eqref{eq: lnL_appendix}, and $6$ in the full model that uses the likelihood in Eq.~\eqref{eq: lnLm_appendix}.

As a consistency check for this decision, we have also produced a set of results using the full likelihood function treating all of the IISM parameters as free parameters. An example set of results is shown in Fig.~\ref{fig:J1853-0004_nomarg}, which uses data from the pulsar J1853-0004, similarly to Fig.~\ref{fig:J1853-0004_spa_corner}, for the sake of comparison. In these results we see the sampled posteriors are not identical to those in the marginalsed case, but have fully consistent confidence intervals and overall behaviour for the noise models. We further see that the posteriors for the annual modulation parameters $\psi^{(c)} = -0.01^{+0.01}_{-0.01}$ and $\psi^{(s)} = 0.00_{-0.01}^{+0.01}$ are consistent with null values, while $\psi^{(0)} = 0.04_{-0.03}^{+0.03}$ and $\psi^{(1)} = 0.23_{-0.03}^{+0.03}$ have non-zero median estimates. The strong anti-correlation between $\psi^{(0)}$ and $\psi^{(2)}$, which is weakly constrained around 0, may indicate a degeneracy between these parameters and that the second-order term is over-fitting the data. Although this is a single example from our population, we observe that the noise models in general throughout our population are not strongly affected by this marginalisation procedure, which suggests that in most pulsars the generic white and red noise models given above are the dominant sources of temporal variability. 

\begin{figure*}[htbp]
    \centering
    \includegraphics[width=0.95\textwidth]{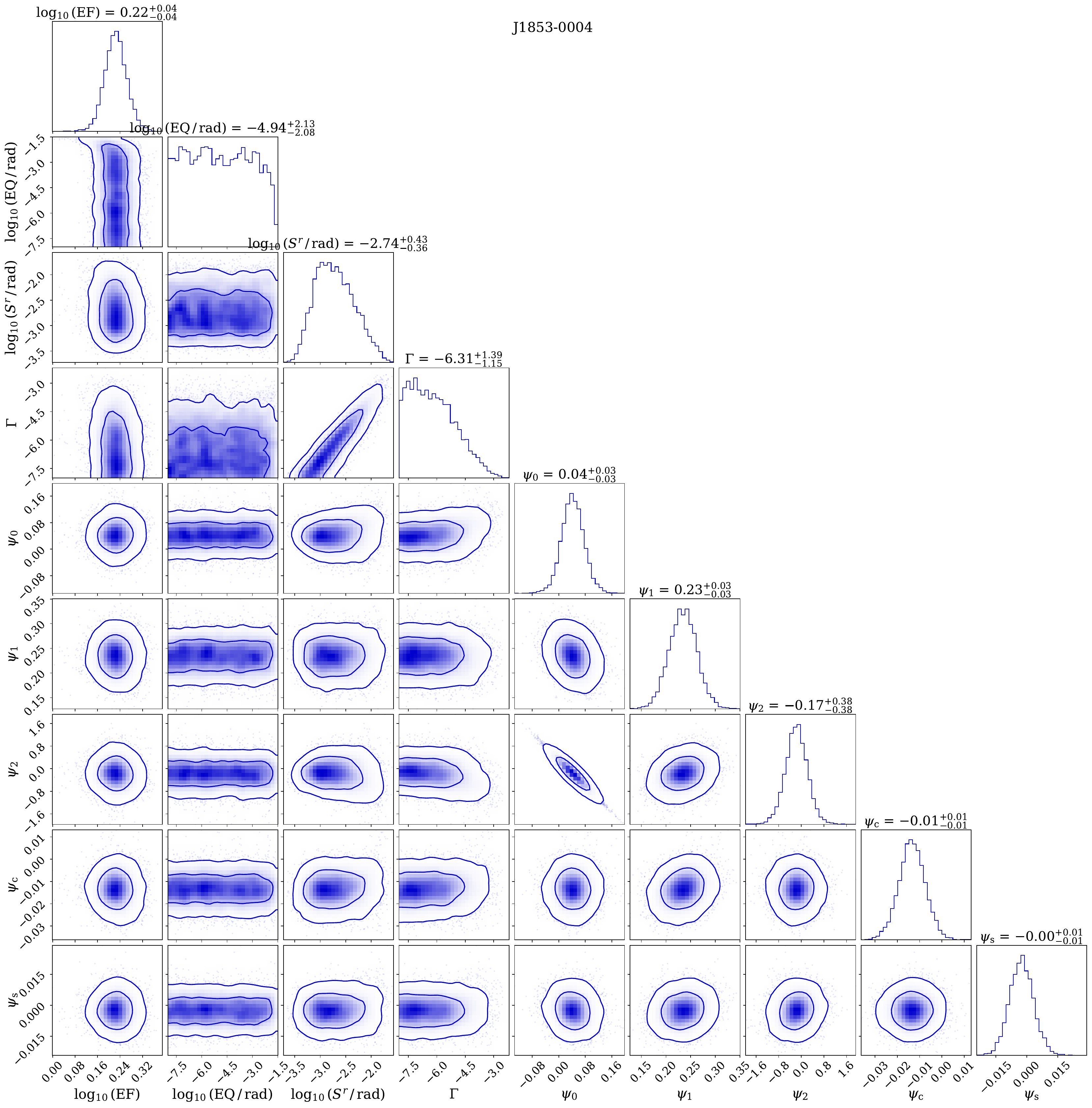}
    \caption{Bayesian posterior probability distributions for the parameters EFAC, EQUAD, $S^{\rm r}$, $\Gamma$, and $\psi^{(i)}$ for the pulsar J1853-0004. This figure is similar to Fig.~\ref{fig:J1853-0004_spa_corner}, but with a full non-marginalised likelihood function calculated from Eq.~\ref{eq: lnL}.}
    \label{fig:J1853-0004_nomarg}
\end{figure*}


\bibliography{ref}

\end{document}